%% file: main.tex
\documentclass[conference,compsoc]{IEEEtran}

\usepackage{url}
\usepackage{hyperref}

\usepackage{tikz}
\usepackage{amsmath}
\usepackage{amssymb}
\usepackage{tabularray}
\usepackage{xspace}
\usepackage{array,graphicx}

\usepackage{multirow}
\usepackage{colortbl}
\usepackage{float}
\usepackage{makecell}
\usepackage{bm}
\usepackage{url}
\usepackage{caption}
\usepackage{subcaption}
\usepackage{anyfontsize}
\usepackage{cite}

\usepackage[most]{tcolorbox}

\usepackage{xcolor}
\definecolor{lightgray}{gray}{0.8}
\definecolor{lightblue}{rgb}{0.55, 0.85, 0.9}

\newtcolorbox{highlightedbox}[1][]{
  colback=lightblue,
  sharp corners,
  boxrule=0pt,
  boxsep=0pt,
  left=1pt,
  right=1pt,
  top=1pt,
  bottom=2pt,
  breakable,
  #1
}

\newcommand{\para}[1]{{\vspace{2pt} \bf \noindent #1 \hspace{1pt}}}

\newcommand{\eg}{e.g.,\xspace}

\newcommand{\ie}{i.e.,\xspace}

\newif\ifshowcomment
\showcommenttrue

\ifshowcomment
    \newcommand{\sifat}[1] {{\footnotesize\color{red}[sifat: #1]}}

    \newcommand{\pgao}[1] {{\footnotesize\color{red}[Peng: #1]}}
    
    \newcommand{\ques}[1] {{\footnotesize\color{red}[ques: #1]}}
    \newcommand{\mj}[1] {{\footnotesize\color{green}[MJ: #1]}}
\else
    \newcommand{\sifat}[1]{}
    \newcommand{\pgao}[1]{}
    \newcommand{\ques}[1]{}
    \newcommand{\mj}[1]{}
\fi

\newcommand{\dct}{{DCT}}
\newcommand{\cnn}{{CNN-F}}
\newcommand{\resyn}{{Resynthesis}}
\newcommand{\texture}{{Gram-Net}}
\newcommand{\meso}{{MesoNet}}
\newcommand{\univclip}{{UnivCLIP}}
\newcommand{\uclip}{{UnivCLIP}}
\newcommand{\patch}{{Patch-Forensics}}
\newcommand{\clsr}{{CLIP-Score}} 
\newcommand{\ularge}{{UnivConv2B}}
\newcommand{\defake}{{DE-FAKE}}

\newenvironment{packed_itemize}{
\begin{list}{\labelitemi}{\leftmargin=1.em}
  \setlength{\itemsep}{3pt}
  \setlength{\parskip}{0pt}
  \setlength{\parsep}{0pt}
  \setlength{\headsep}{0pt}
  \setlength{\topskip}{0pt}
  \setlength{\topmargin}{0pt}
  \setlength{\topsep}{0pt}
  \setlength{\partopsep}{0pt}
}{\end{list}}

\newcolumntype{"}{!{\vrule width 1pt}}

\makeatletter
\def\hlinewd#1{%
\noalign{\ifnum0=`}\fi\hrule \@height #1 \futurelet
\reserved@a\@xhline}
\makeatother

\begin{document}

\title{An Analysis of Recent Advances in Deepfake Image Detection in an\\ Evolving Threat Landscape}

\author{\IEEEauthorblockN{Sifat Muhammad Abdullah, 
Aravind Cheruvu, Shravya Kanchi,\\ Taejoong Chung, Peng Gao, Murtuza Jadliwala\IEEEauthorrefmark{1}, Bimal Viswanath}

Virginia Tech, \IEEEauthorrefmark{1}UT San Antonio%
\\

\{sifat, acheruvu, shravya, tijay, penggao, vbimal\}@vt.edu, \IEEEauthorrefmark{1}murtuza.jadliwala@utsa.edu
}

\maketitle

\input{abstract}

\begin{IEEEkeywords}
deepfake image, foundation models, generative models, deepfake detection
\end{IEEEkeywords}

\input{intro}

\input{background}

\input{baselines}

\input{evaluation1}

\input{evaluation2}

\input{evaluation3}

\input{related-work}

\input{conclusion}

\bibliographystyle{IEEEtran}
\bibliography{main.bib}

\input{appendix}

\end{document}

%% file: abstract.tex
\begin{abstract}

Deepfake or synthetic images produced using deep generative models pose serious risks to online platforms. This has triggered several research efforts to accurately detect deepfake images, achieving excellent performance on publicly available deepfake datasets. In this work, we study 8 state-of-the-art detectors and argue that they are far from being ready for deployment due to two recent developments. First, the emergence of lightweight methods to customize large generative models, can enable an attacker to create many customized generators (to create deepfakes), thereby substantially increasing the threat surface. We show that existing defenses fail to generalize well to such \emph{user-customized generative models} that are publicly available today. We discuss new machine learning approaches based on content-agnostic features, and ensemble modeling to improve generalization performance against user-customized models. Second, the emergence of \textit{vision foundation models}---machine learning models trained on broad data that can be easily adapted to several downstream tasks---can be misused by attackers to craft adversarial deepfakes that can evade existing defenses. We propose a simple adversarial attack that leverages existing foundation models to craft adversarial samples \textit{without adding any adversarial noise}, through careful semantic manipulation of the image content. We highlight the vulnerabilities of several defenses against our attack, and explore directions leveraging advanced foundation models and adversarial training to defend against this new threat.

\end{abstract}

%% file: intro.tex
\section{Introduction}
\label{sec:intro}

\noindent Recent advances and applications of generative AI have catapulted this technology as the next frontier in Artificial Intelligence~\cite{Generati25:online}. Generative models are a family of machine learning (ML) algorithms capable of learning a data distribution to produce new \textit{synthetic} variations of these data. Generative models can produce convincing synthetic images, which can be easily misused, raising several security threats. Easily available, off-the-shelf generative models (\eg Stable Diffusion~\cite{rombach2022high}, DALL·E~\cite{ramesh2021zero}, StyleCLIP~\cite{patashnik2021styleclip}) can be used to create synthetic or \textit{deepfake} imagery to power large-scale fake account campaigns on social media platforms~\cite{Thelates91:online}, create media for convincing fake news articles~\cite{AIgenera60:online}, create fake pornographic images~\cite{cnndeepfakevdeio}, spoof identity verification in financial services~\cite{Liveness-news, li2022seeing}, and power other threats. Countries across the globe are struggling to respond to the risks posed by generative AI, as false alarms raised by poorly implemented defenses can completely erode our trust in online content~\cite{AsDeepfa1:online}.

The urgency of this problem triggered a flurry of research efforts 
that proposed methods to detect deepfake images~\cite{ojha2023towards, ricker2022towards, liu2020global, he2021beyond, wang2020cnn, afchar2018mesonet, chai2020makes}. State-of-the-art (SOTA) detection schemes use a supervised learning scheme that leverages ``imperfections'' in fake images, to distinguish fake from real images. They do so using a variety of methods (Section~\ref{sec:defense_criteria}), \eg using texture statistics~\cite{liu2020global}, finding imperfections in the frequency spectrum~\cite{ricker2022towards} or local patches~\cite{chai2020makes}. All these defenses claim extremely high detection accuracy on the datasets they were evaluated on (Section~\ref{sec:defense_criteria}).

\para{New threat vectors.}
In this work, we argue that \textit{these defenses face a rapidly evolving threat landscape}, placing them at severe risk of underperforming in the real world. This evolving threat landscape is fueled by two recent advances in machine learning: 

\begin{packed_itemize}
\item \textbf{Emergence of lightweight methods that allow users to customize large generative models}, thereby enabling democratization of generative AI technologies. Prior defenses were primarily evaluated using images from a few instances of generative models from different families, mostly GANs~\cite{karras2019style, brock2018large} and Diffusion models~\cite{rombach2022high}. Today, the threat landscape has changed dramatically---\eg there are over 3,000 user-customized\footnote{As of Feb, 2024} (\ie by Internet users) variants of the Stable Diffusion~\cite{rombach2022high} model alone on platforms like CivitAI~\cite{CivitaiT97:online} and Huggingface~\cite{ModelsHu5:online}. This is enabled by new algorithmic methods that enable efficient fine-tuning of these large generative models in resource-constrained setups (limited training data and compute power)~\cite{hu2021lora}. Previous defense efforts have documented the challenges of achieving high generalization performance across a few generative models and families~\cite{ricker2022towards}.
The current threat landscape that contains thousands of publicly available variants of generative models, where any can be misused, presents an unprecedented and challenging environment for defenses.

\item \textbf{Emergence of vision foundation models.} Foundation models are ML models trained on broad data (usually using self-supervision), which can be further adapted for a variety of downstream tasks with impressive performance~\cite{bommasani2021opportunities}. Popular vision foundation models include CLIP-ResNet~\cite{radford2021learning}, EfficientNet~\cite{tan2019efficientnet} and ViT~\cite{dosovitskiy2020vit}. 
For example, 
the CLIP model can learn a joint embedding space for both text and image inputs and can function as a generic text and image encoder. It can then be further adapted to build a variety of downstream tasks, \eg to build computer vision classifiers. We show that such foundation models can be easily integrated with existing generative models to craft \textit{``adversarial fake images''}, \ie fake images that can fool deepfake classifiers.
\end{packed_itemize}

\begin{figure}[t]
\centering
\begin{subfigure}[t]{0.4\columnwidth}
\centering 
\includegraphics[height=0.75\textwidth,width=0.75\columnwidth]{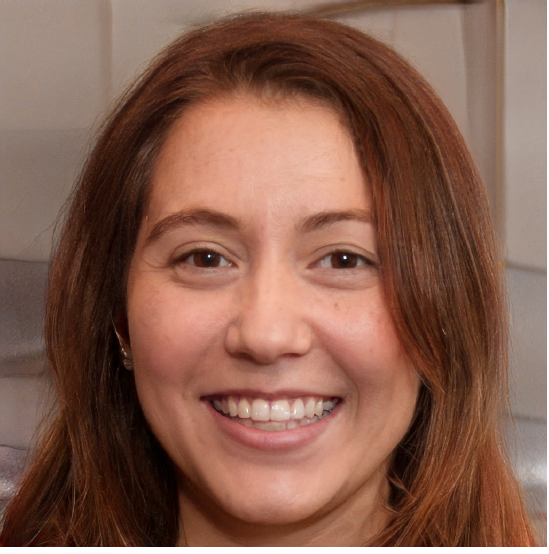}
\caption{Source image}
\end{subfigure}
\begin{subfigure}[t]{0.4\columnwidth}
\centering 
\includegraphics[height=0.75\textwidth,width=0.75\columnwidth]{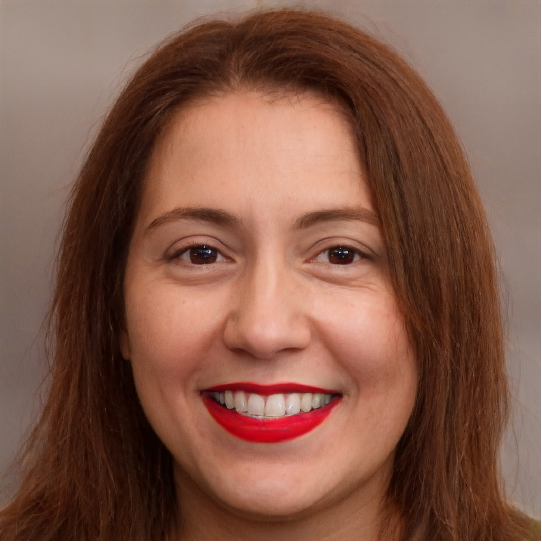}
\caption{Manipulated image}
\end{subfigure}
\caption{Adding lipstick in the manipulated image evades a deepfake detector~\cite{ricker2022towards}.}
\vspace{-0.1in}
\label{fig:evasive-df-sample}
\end{figure}

\para{Contributions.}
We conduct the first large-scale study analyzing 8 SOTA deepfake defenses by considering the above evolving threats. Our key contributions are as follows:

\begin{packed_itemize}
\item \textit{We provide a critique of the methods used to train and evaluate existing defenses (Section~\ref{sec:defenselimitations})}. We tried to reproduce the findings of 8 state-of-the-art defenses, and in the process identified several issues related to the training and evaluation methodologies used in these works. For example, the most recent defense, \uclip{}~\cite{ojha2023towards} does not control the content and quality of images used in the training and evaluation dataset, leading to possible spurious correlations being learned. This can result in overestimating the performance of these defenses, resulting in misplaced confidence in the strengths of these defenses in the real world. We hope to educate the community about these issues and suggest actionable steps to correctly train and evaluate defenses.

\item \textit{We study defense effectiveness in a threat landscape enabled by the democratization of generative AI technologies (Section~\ref{sec:generalization_testing}).} We study such a setting by focusing on the many publicly available, user-customized variants of the Stable Diffusion model, particularly 16 of them. 
All defenses exhibit significant degradation in performance when applied to these user-customized variants, on average up to 53.92\% degradation in Recall (over the 16 models). Two notable recent defenses, \uclip{} and \defake{}~\cite{sha2023fake} that are state-of-the-art in terms of generalization performance, also show significant degradation in performance.
We present new strategies to further improve generalization performance. This includes augmenting existing defenses with \textit{content-agnostic} features, and using ensemble models that combine defenses using foundation model features and domain-specific features (\eg{} frequency-based features~\cite{ricker2022towards}).

\item \textit{We study defenses against an adversary who leverages vision foundation models to create adversarial samples \underline{without} adding adversarial noise (Section~\ref{sec:advattack}).} It is important to study the adversarial robustness of deepfake defenses. Existing work has focused mainly on adding adversarial noise (perturbations) to fake images to evade detection~\cite{hou2023evading, carlini2020evading}. This can degrade image quality, especially when they appear in regions with a smooth texture~\cite{li2021exploring}.
\textit{We show that foundation models can be leveraged to successfully evade defenses without adding any noise to the images.} Our key idea is to leverage a foundation model to create an adversarial fake image by making careful \textit{semantic changes} to the image content. 
For example, Figure~\ref{fig:evasive-df-sample} shows a successful attack, where manipulating the lip color is sufficient to fool an existing detector~\cite{ricker2022towards}. To achieve this, the attacker uses a text prompt (that describes the semantic change) on an adversarially updated image generator. The image generator is adversarially updated using a \textit{surrogate deepfake classifier powered by a foundation model}.
Our attack can significantly degrade the performance of all 8 defenses. We identify defenses that are notably weak and resilient against our attacks---defenses using frequency-based features are highly vulnerable to our attack, while defenses using a foundation model themselves are more resilient. We also explore two strategies to improve adversarial resilience: (1) defenses that use more powerful foundation models (\ie pretrained on larger datasets), compared to the foundation model used by the attacker, demonstrate more resilience, and (2) adversarial training can be an effective temporary measure to build resilience.

\end{packed_itemize}

Our contributions highlight the urgent need to rethink defenses  in a setting where the adversary can customize and create their own deepfake generators and incorporate powerful foundation models to create evasive deepfakes. Section~\ref{sec:future-work} discusses several directions for future work. 
\textit{Code and data used in this study are available on Github.\footnote{\href{https://github.com/secml-lab-vt/EvolvingThreat-DeepfakeImageDetect}{github.com/secml-lab-vt/EvolvingThreat-DeepfakeImageDetect}}}

%% file: background.tex

\section{Background and Threat Model} 
\label{sec:background}

\para{Deepfakes and real images.} We use the term ``deepfakes'' or ``fake images'' to refer to \textit{fully synthetic images} created using \textit{generative models}. The deepfake images can have any type of content, \ie we \textit{do not} place restrictions on the type of content such as only faces.
Partially synthetic deepfakes, such as face-swaps or face-reenactments are not considered in this work~\cite{mirsky2021creation}. Today, state-of-the-art generative models for images are based on GANs~\cite{kang2023scaling}, Variational Autoencoders (VAEs)~\cite{van2017neural}, and Diffusion models~\cite{rombach2022high, saharia2022photorealistic}. Recently, the integration of multimodal vision-language models into image generation pipelines has enabled individuals to create an image by just supplying a prompt that describes the desired content~\cite{radford2021learning, cherti2023reproducible, rombach2022high, ramesh2022hierarchical}. This increased ease of creating synthetic content can be misused to create misleading content.

We use the term ``real image'' to refer to any image that is not produced by a generative model. This includes images produced through photography, human-made digital, or any content created by humans. Note that prior work primarily considered camera images to be real images. In contrast, our definition encompasses a much wider class of images in the real set. Given the capabilities of generative models to create digital art and other types of content~\cite{Midjourn13:online}, it is important to broaden the definition.

\para{Foundation models.} In this work, we use foundation models as general-purpose feature extractors. These models are usually trained on Internet-scale or large datasets of image and text modalities, mostly using self-supervised learning strategies---a process known as pretraining. These models learn highly generalizable representations of the different input modalities, \eg text or images, making them highly adaptable for various downstream tasks, \eg image or text classification. An image classifier can be built using the features extracted from a foundation model. For example, ViT~\cite{dosovitskiy2020vit} is a transformer-based foundation model trained on 14M images from the ImageNet-21K~\cite{deng2009imagenet} dataset, and achieves excellent performance on challenging tasks, \eg the VTAB~\cite{zhai2019large} suite of 19 tasks such as image classification, object detection, and localization. We used multiple vision foundation models, namely, EfficientNet~\cite{tan2019efficientnet}, ViT, and models from the CLIP and OpenCLIP family~\cite{radford2021learning, cherti2023reproducible}. Due to their pretraining, we can extract highly effective features to build deepfake classifiers.

\para{Threat model.}
The attacker uses a generative model to create convincing and high-quality deepfake images that capture a desired target content. Any type of content can be generated. To enable the generation of desired content, we consider the use of text-to-image generative models, namely Stable Diffusion (SD)~\cite{rombach2022high} and StyleCLIP~\cite{patashnik2021styleclip}. For SD, the attacker starts with only a text prompt that describes the desired content. For StyleCLIP, the attacker uses both a text prompt and a source image. StyleCLIP translates the source image into a target image that captures the content described in the text prompt.

The defender aims to distinguish fake images from real images using a supervised machine learning (ML) model. We consider 8 publicly available state-of-the-art (SOTA) ML schemes (in the research literature) to detect fake images. In Section~\ref{sec:advattack}, we further consider a defender who is aware of the generative model used by the attacker and optimizes the defense to detect images from that generative model.

The attacker may adapt the generator to produce \textit{adversarial} fake images that can fool detectors, while preserving the desired quality and content of the image. This is done \emph{without} adding any adversarial noise to the generated image, \ie the generated image itself is adversarial by design. Instead of adding adversarial noise, the attacker makes adversarial semantic changes to the content, using a text prompt and an adversarially updated image generator. We consider a full black-box setting where the attacker has no query access or access to the defender's detection model.

%% file: baselines.tex
\section{Generative Models and Defenses}
\label{sec:gen_models_and_defenses}
\subsection{Generative Models to Create Deepfakes}
\label{sec:only_gen_models} 
\noindent Generative models learn the underlying patterns in the training data to generate novel content. We focus on two popular generative models ---Stable Diffusion (SD)~\cite{rombach2022high} and StyleCLIP~\cite{patashnik2021styleclip}. Both models are capable of producing high-quality imagery, are open-source, and the pretrained models are publicly available, allowing us to study different attacks.

 \para{Stable Diffusion (SD).} We use the SD model to study the impact of democratization of AI technologies on defense efforts. The SD model is widely popular, with users ``customizing'' or fine-tuning this model on datasets to further improve image quality or adapt to newer data distributions and share the models publicly. We see over $3,000$ user-customized variants of these models being shared on CivitAI~\cite{CivitaiT97:online} and HuggingFace~\cite{ModelsHu5:online}.

SD is implemented as a text-to-image generation model based on the Latent Diffusion Model (LDM)~\cite{rombach2022high}. At its core, SD acts as a denoiser. Starting from a noise vector (\eg Gaussian noise), SD can transform it into a complex target distribution (an image) conditioned on text prompts through a series of invertible operations to generate high-quality images. To reduce the computational demands, SD implements this diffusion process in a low-dimensional latent space, instead of the pixel space. The input text prompt is encoded using the CLIP~\cite{radford2021learning} text encoder, which has learned a joint language-image embedding space.

\para{StyleCLIP.} We use the StyleCLIP model to study the adversarial robustness of existing defenses. StyleCLIP is a text-driven image modification model, \ie given a source image and a text prompt describing the target content, StyleCLIP manipulates the source image to capture the desired target content. For example, given a face image, StyleCLIP can manipulate facial attributes (\eg hair, eyes). StyleCLIP uses StyleGAN2~\cite{karras2020analyzing} (GAN family) as the image generator, 
and the OpenAI CLIP model~\cite{radford2021learning} to perform text-driven image modifications. Generative Adversarial Networks (GANs) have been considered state-of-the-art for image generation for almost a decade~\cite{goodfellow2020generative, kang2023scaling}. A GAN model includes a generator and a discriminator, which are trained adversarially. The generator aims to generate fake images that can fool the discriminator, and the discriminator's classification feedback is used by the generator to improve its quality of image generation. Being a GAN model, it is a perfect fit for our setting---we adversarially update the StyleGAN2 generator of StyleCLIP to create adversarial fake images.

Image modifications are enabled by manipulations to an intermediate latent space in StyleGAN2. StyleCLIP uses the intermediate latent space called Stylespace S. Each stylespace vector contains channels that are disentangled latent representations of the color and semantics of an input image. 
Using the target text prompt, encoded to a joint text and image embedding space, StyleCLIP infers a direction in the stylespace to drive the manipulations, \ie identifies which channels in the stylespace should be manipulated to satisfy the target text prompt. There are two key parameters at generation time: (1) $\beta$: stylespace channels with relevance score higher than $\beta$ are manipulated. (2) $\alpha$: controls the strength of manipulations made to a stylespace channel.

\para{Other models.} We considered other generative models but did not include them for one or more of the following reasons: unavailability of training code or pretrained checkpoints and poor quality imagery. Details are in Appendix~\ref{appendix:defense-implementation}.

\subsection{Defenses: Deepfake Detection Schemes}
\label{sec:defense_criteria}

\noindent We select 8 supervised learning-based defenses using the following criteria: \textit{(1) Performance:} All 8 defenses claim impressive detection performance and were examined in previous work on deepfake defenses. \textit{(2) Availability:} The availability of model checkpoints and training code is a requirement for our methodology, as we need to fine-tune these defenses on different datasets. \textit{(3) Target deepfakes:} We only study defenses designed to detect fully synthetic images. Defenses made to detect partially synthetic content (\eg face-swapped content~\cite{xu2022mobilefaceswap}) are not our focus.
\textit{(4) Content types:} Previous work mainly focused on detecting face deepfakes. We do not place any content restrictions, given the emerging threat that any arbitrary content can be a deepfake. In addition to face images, we consider several content types, \eg artwork, illustrations, images of different objects. \textit{(5) Diverse methodologies:} The chosen defenses use diverse methodologies, thus helping to understand the robustness of different defense strategies.

\para{\univclip{}~\cite{ojha2023towards}.} This is the most recent defense (in 2023) with two key highlights: \textit{First}, \uclip{} is one of the first defenses that uses a large foundation model to build a deepfake detector. The CLIP:ViT-L/14 foundation model~\cite{radford2021learning}, trained on 400M image-text pairs, is used. After extracting features from the \textit{(frozen)} CLIP:ViT model,
the study recommends using either a nearest neighbor classifier or a linear classification layer, with further training to predict an image as real or fake.
We use the linear classifier approach as it performs better in our setting (Section~\ref{sec:defenselimitations}). \textit{Second}, authors claim that extracting features from a foundation model, which has not been explicitly trained for a deepfake detection task, provides (surprisingly) high generalization performance. \uclip{} is shown to achieve up to 99.17\% Average Precision in generalizing to fake images from (previously unseen during training) generative models.

\para{\defake{}~\cite{sha2023fake}.} Similar to \uclip{}, Sha et al.~\cite{sha2023fake} also use the CLIP~\cite{radford2021learning} model to build a detector. Compared to \uclip{}, a key difference is to augment the image's embedding along with an embedding of the text prompt (both extracted using CLIP) to train the detector.
The intuition is that real images usually have more information than their respective captions, whereas fake images generated from prompts only show content that is specific to that prompt. This disparity in information is used to detect deepfakes. The classifier can effectively generalize to deepfakes from models not seen during training, \eg achieving an accuracy of 90.9\% on DALL·E 2 images. As recommended by the authors, we use the image captioning model BLIP~\cite{li2022blip} to generate the prompts for training. This fits our threat model as we consider a defender who will only receive an image for detection, \ie without an associated prompt.

\para{\dct{}~\cite{ricker2022towards}.} Ricker et al.~\cite{ricker2022towards} show that the frequency domain provides discriminatory features for deepfake detection. This is inspired by previous work showing visible artifacts, \eg grid-like patterns, in the frequency spectrum of GAN-generated images~\cite{frank2020leveraging}. To build the classifier, the frequency domain features are extracted from the images using a discrete cosine transform (DCT). We use the log-scaled version of the DCT features as recommended by Rocker et al.~\cite{ricker2022towards} for improved performance. These features are used to train a Logistic Regression classifier. DCT achieves 97.7\% and 73\% accuracy on images generated by GAN and Diffusion model, respectively.

\para{\patch{}~\cite{chai2020makes}.} This detector is designed only for face content and claims high generalization performance across generative models. The key intuition is that searching for artifacts in local patches of the image provides more generalizable patterns for detection, compared to looking for global artifacts (\ie in the image as a whole). To identify local artifacts, an image is broken down into equal sized patches and a patch-based classifier is trained. This classifier uses a truncated convolutional neural network with small receptive fields that are better suited for identifying local imperfections. Classification decisions at the patch level are aggregated into an overall prediction. We use the Xception Block 2 variant of their patch-based classifier, which demonstrated impressive performance. \patch{} is shown to achieve 100\% Average Precision when applied to fake images from StyleGAN.

\para{\texture{}~\cite{liu2020global}.} This scheme has been extensively studied to detect fake faces, but also claims to generalize to other content. The key insight is that the texture statistics of fake images (\eg face content) are significantly different from real images. Using global texture features was found to be robust against image distortions, and also generalize across different GANs. Based on this observation, the authors propose a novel CNN-based architecture called ``Gram-Net'' that can extract global texture features for detection. We build on a version of their model trained on StyleGAN (fake) and CelebA-HQ (real) images, as this model was used to detect arbitrary deepfake content (\ie not just faces). \texture{} achieves 89.26\% average accuracy when applied to images from generative models not seen during training.

\para{\resyn{}~\cite{he2021beyond}.} This detection scheme also aims to generalize across different generative models. Instead of relying on low-level artifacts specific to certain generative models, this scheme aims to learn more generalizable features for detection. This is done by resynthesizing testing images (\ie both real and fake) based on different auxiliary tasks, \eg super-resolution, denoising and colorization. The resynthesis component is only trained on real images, and therefore results in different residuals (reconstruction errors) for fake images. The key insight is that the residuals from the resynthesis tasks provide effective discriminatory features. The detector is trained jointly with the synthesizer and is shown to achieve 93.7\% average accuracy.

\para{\cnn{}~\cite{wang2020cnn}.} This is a widely studied defense. The key idea is that CNN-based generators leave detectable \textit{fingerprints}. A ResNet-50 model (pretrained on ImageNet~\cite{krizhevsky2012imagenet}) is fine-tuned to learn such fingerprints. A notable step of training is a data augmentation strategy based on standard post-processing schemes that shows improved generalization performance. The work highlights that the detector needs to be trained only on images from a single CNN-based generator to generalize across different fake sources. \cnn{} is shown to achieve 93\% mean Average Precision.

\para{\meso{}~\cite{afchar2018mesonet}.} Originally designed to detect deepfake videos, this scheme also functions as a deepfake image detector. The key idea is to conduct analysis at a \textit{mesoscopic level}. At a macroscopic level, it is hard to identify any semantic differences between fake and real images. The authors claim that a microscopic analysis of artifacts can be challenging as well, and hence they focus on mesoscopic properties. They propose a unique DNN with a small number of layers to extract mesoscopic features. They also note that replacing convolution layers with Inception modules~\cite{szegedy2015going} leads to better classification results. \meso{} claims to achieve 98.4\% accuracy.

\subsection{Defense Implementations}
\label{sec:defense-implementation}

\noindent To effectively evaluate the threats, we need to consider a capable defender. Existing defense implementations (pretrained models, whenever available) are trained on different datasets, thus generalizing differently to newer datasets. We need to implement (train) each defense to have the best chance of detection in our threat settings. Thus, we resort to retraining all 8 defenses on 2 datasets relevant to our research goals. \textit{For each dataset, we strive to use images with highest visual quality (for fake and real), and also ensure a similar content distribution for fake and real classes---so the classifiers can learn effective features and not learn to separate based on differences in semantic content.}

\begin{figure}[t]
\centering 
\begin{subfigure}[t]{0.48\columnwidth}
\centering 
\includegraphics[height=0.99\textwidth,width=0.7\columnwidth]{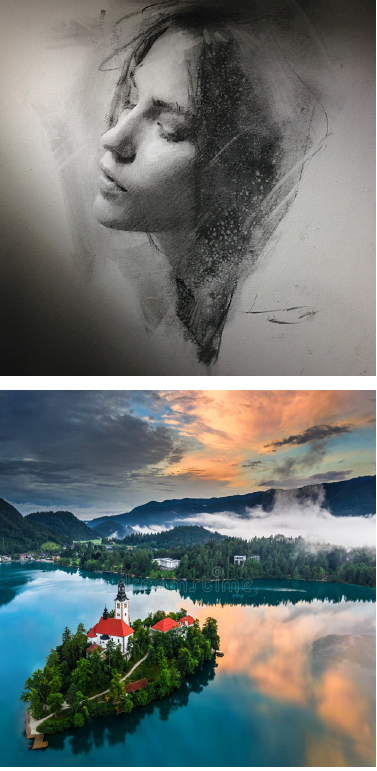}
\caption{Real}
\end{subfigure}
\begin{subfigure}[t]{0.48\columnwidth}
\centering 
\includegraphics[height=0.99\textwidth,width=0.7\columnwidth]{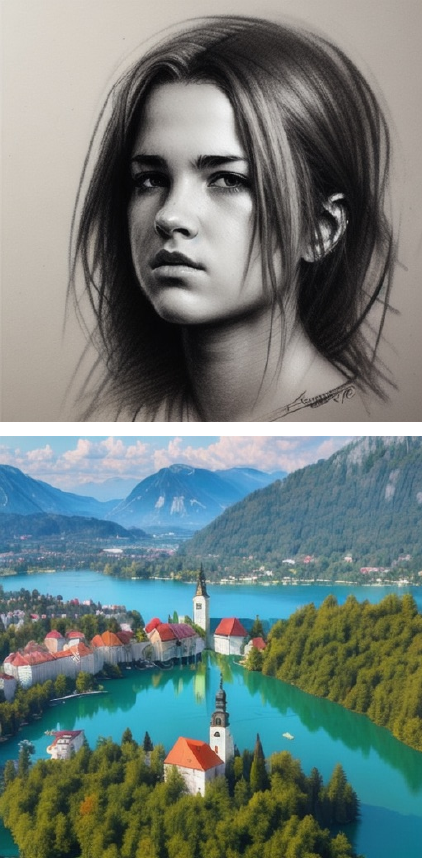}
\caption{Fake}
\end{subfigure}  
\caption{Real and fake samples from our SD dataset.}
\label{fig:realfakecomparison}
\end{figure}

\para{Training/evaluation dataset 1: SD dataset.} This dataset is used to investigate defense effectiveness on user-customized generative models (Section~\ref{sec:generalization_testing}). \textit{Real images} are sampled from the LAION-AESTHETICS dataset~\cite{LAIONAes38:online}. This dataset contains 625K image-text pairs of highest visual quality, as rated by the LAION-Aesthetics Predictor V2~\cite{christop57:online} model. \textit{The images cover a wide variety of content, \eg people, nature, objects, illustrations, and digital art.} Fake images are generated with a square aspect ratio. So, we further filter to extract image-text pairs, where the width/height of the images are roughly 500 pixels (with an error margin of 150 pixels), and also remove images flagged as unsafe (based on available metadata). Finally, we randomly sample a subset from this filtered set. \textit{Fake images} are created using the Realistic Vision v1.4 SD model~\cite{SG1612223:online}, with text prompts obtained from the real dataset (comes as image-text pairs). This ensures similarity in content between the fake and real classes. The Realistic Vision model was created by fine-tuning the SDv1.5 model to enhance image quality. It is widely used with over half a million downloads. We use this SD model instead of the base SD models (\eg SDv1.5~\cite{runwayml61:online}) because of its ability to generate higher quality, realistic, aesthetic images. Images are generated as 512$\times$512 using the default settings. Figure~\ref{fig:realfakecomparison} shows examples of fake and real images from this dataset. In total, we collected a \textit{balanced} dataset of $16K$, $2K$, and $2K$ images across both classes for training, validation, and testing, respectively.

\begin{table}[!t]
\centering
\small
\setlength{\tabcolsep}{1.5pt}
\setlength\extrarowheight{3pt}
\begin{tabular}{l"ccc"ccc}
\multirow{2}{*}{\textbf{Defense}} & \multicolumn{3}{c"}{\textbf{SD}} & \multicolumn{3}{c}{\textbf{StyleCLIP}} \\ \cline{2-7} 
 & \multicolumn{1}{c|}{\textbf{Precision}} & \multicolumn{1}{c|}{\textbf{Recall}} & \textbf{F1} & \multicolumn{1}{c|}{\textbf{Precision}} & \multicolumn{1}{c|}{\textbf{Recall}} & \textbf{F1} \\ \hlinewd{1.1pt}
\textbf{\uclip{}} & \multicolumn{1}{c|}{90.20} & \multicolumn{1}{c|}{93.90} & 92.01 & \multicolumn{1}{c|}{93.79} & \multicolumn{1}{c|}{92.20} & 92.99 \\ 
\textbf{\defake{}} & \multicolumn{1}{c|}{93.82} & \multicolumn{1}{c|}{94.20} & 94.01 & \multicolumn{1}{c|}{74.41} & \multicolumn{1}{c|}{78.80} & 76.54 \\ 
\textbf{\dct{}} & \multicolumn{1}{c|}{100} & \multicolumn{1}{c|}{88.80} & 94.07 & \multicolumn{1}{c|}{100} & \multicolumn{1}{c|}{99.60} & 99.80 \\ 
\textbf{\patch{}} & \multicolumn{1}{c|}{-} & \multicolumn{1}{c|}{-} & - & \multicolumn{1}{c|}{91.76} & \multicolumn{1}{c|}{91.30} & 91.53 \\ 
\textbf{\texture{}} & \multicolumn{1}{c|}{99.99} & \multicolumn{1}{c|}{99.10} & 99.55 & \multicolumn{1}{c|}{99.99} & \multicolumn{1}{c|}{99.60} & 99.80 \\ 
\textbf{\resyn{}} & \multicolumn{1}{c|}{85.39} & \multicolumn{1}{c|}{86.50} & 85.94 & \multicolumn{1}{c|}{98.80} & \multicolumn{1}{c|}{98.70} & 98.75 \\ 
\textbf{\cnn{}} & \multicolumn{1}{c|}{99.41} & \multicolumn{1}{c|}{83.80} & 90.94 & \multicolumn{1}{c|}{99.90} & \multicolumn{1}{c|}{97.10} & 98.48 \\ 
\textbf{\meso{}} & \multicolumn{1}{c|}{99.99} & \multicolumn{1}{c|}{98} & 98.98 & \multicolumn{1}{c|}{96.70} & \multicolumn{1}{c|}{99.50} & 98.08
\end{tabular}%
\caption{Performance of defenses, reported as Precision, Recall and F1 scores (\textbf{in percentage}) for the fake class of test set, after our training / fine-tuning. \patch{} is marked as `-' as it is only applicable to face content, and therefore not evaluated on SD images.}
\label{tab:before_after_defense_test}
\end{table}

\para{Training/evaluation dataset 2: StyleCLIP dataset.} This dataset is used to train defenses to study robustness against an adversary leveraging vision foundation models (Section~\ref{sec:advattack}). The attacker uses the StyleCLIP generator which is based on StyleGAN2. We only use images with face content, as StyleCLIP only produces limited content types and is widely used for face content. \textit{Real images} are sampled from the Flickr-Faces-HQ (FFHQ) dataset\cite{karras2019style}, which has $70K$ high-quality, high-resolution (1024$\times$1024) face images. We randomly sample a subset from this dataset. \textit{Fake images} are randomly sampled from the official repository of StyleGAN2 generated images. We use fake images generated using the truncation parameter $\psi=1.0$ which ensures maximum diversity in face images. In total, we collected a balanced set of $16K$, $2K$ and $2K$ images across both classes for training, validation, and testing, respectively.

\para{Training configuration.} We made extensive efforts to limit overfitting and create high-performing versions of each defense. Two versions were developed for each defense, with training conducted on the SD and SyleCLIP datasets, respectively. For all defenses, except \dct{}, we fine-tuned the models starting from the checkpoints provided by the model creators. The \dct{} model is not a good fit for fine-tuning, as it uses a simple logistic regression classifier, plus the authors did not provide a pretrained model. For 6 of the 8 defenses, the hyperparameters recommended by the model creators did not yield high enough validation performance. For these 6 defenses, we further tuned the hyperparameters for optimal performance. Appendix~\ref{appendix:defense-implementation} provides more details on training configurations.

Table~\ref{tab:before_after_defense_test} shows the detection performance of defenses after our training/fine-tuning on our two datasets. We report the F1 score for the fake class. \emph{Note the high F1 scores in most cases.} \patch{} is only applicable to face content and is therefore omitted for evaluation on the SD dataset.

%% file: evaluation1.tex
\section{A Critique of Existing Defense Efforts}
\label{sec:defenselimitations}

\noindent We highlight three key limitations of existing defenses.

\begin{figure}[!t]
\centering
\begin{subfigure}[t]{0.45\columnwidth}
\includegraphics[height=0.8\textwidth,width=\columnwidth]{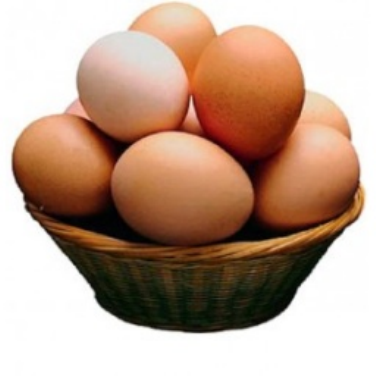}
\caption{Real}
\end{subfigure}
\hfill
\begin{subfigure}[t]{0.45\columnwidth}
\includegraphics[height=0.8\textwidth,width=\columnwidth]{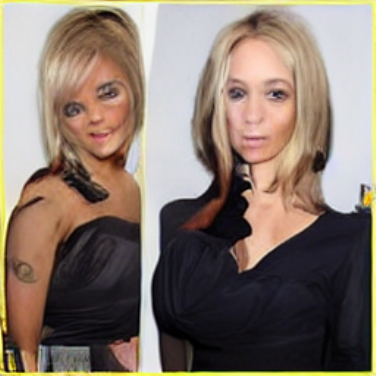}
\caption{Fake}
\end{subfigure}
\caption{Real and fake images used by the \univclip{} defense. Note the poor visual quality of the fake sample.}
\label{fig:univclip_poor}
\end{figure}

\para{Limitation 1: Lack of control for content and image quality.} 
\textit{Controlling content: }Training datasets should have a similar content distribution between the fake and real classes. 
For example, suppose that the real class consists of only car images and fake class consists of only face images. In this case, the model can learn spurious features that are not relevant to the task of distinguishing real and fake images. The model can learn to differentiate between cars and faces, rather than learning features that indicate authenticity of the images.
Given the availability of text-to-image models, content can be controlled by using the caption of a real image as a prompt to generate a fake image with similar content. We find that the most recent defense, \uclip{}, which is SOTA in terms of generalization performance, does not appear to control the content distribution (\ie their training methodology does not discuss this aspect). This is problematic. All other defenses make some effort to control the distribution of content across the two classes. 

\textit{Image quality: }The common premise of the threat of deepfakes is that fake images can appear convincingly real and therefore mislead viewers. Therefore, we encourage the focus on training and evaluating high-quality fake (and real) imagery. Note that training a detector on a dataset where the real images are of high visual quality \emph{but} the fake images are of low quality can lead to a classifier that may not detect deepfakes of high quality in the wild. We observe this for \uclip{}, where the fake images used for training and evaluation are of relatively lower visual quality compared to the real images. Figure~\ref{fig:univclip_poor} shows an example of fake and real images used by \uclip{}. This fake sample was generated using the LDM~\cite{rombach2022high} model and is of low visual quality.

\begin{figure}[t]
\centering
\begin{subfigure}[t]{0.49\columnwidth}
\includegraphics[height=0.565\textwidth,width=1.01\columnwidth]{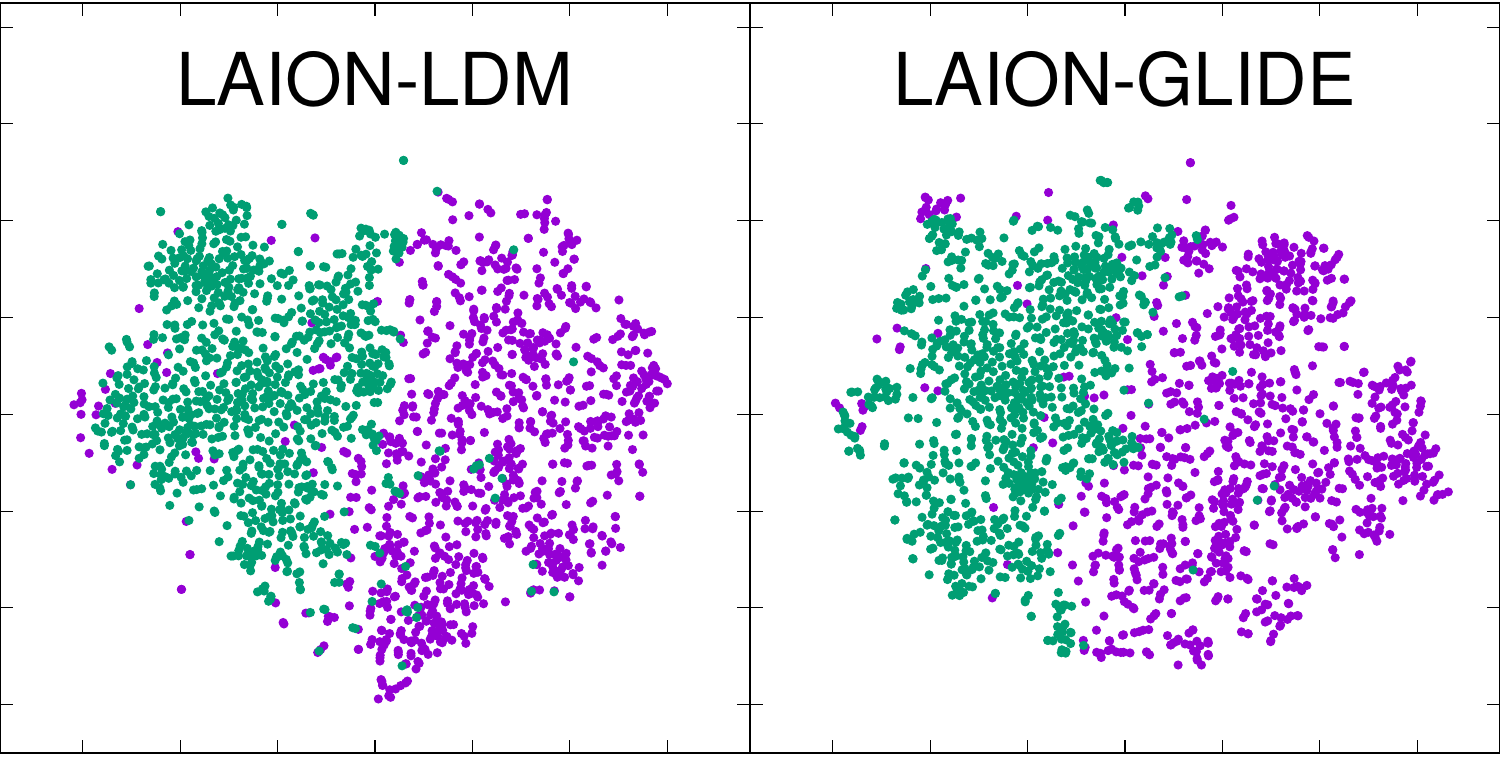}
\caption{UnivCLIP dataset}
\end{subfigure}
\hfill
\begin{subfigure}[t]{0.49\columnwidth}
\includegraphics[height=0.625\textwidth,width=1.01\columnwidth]{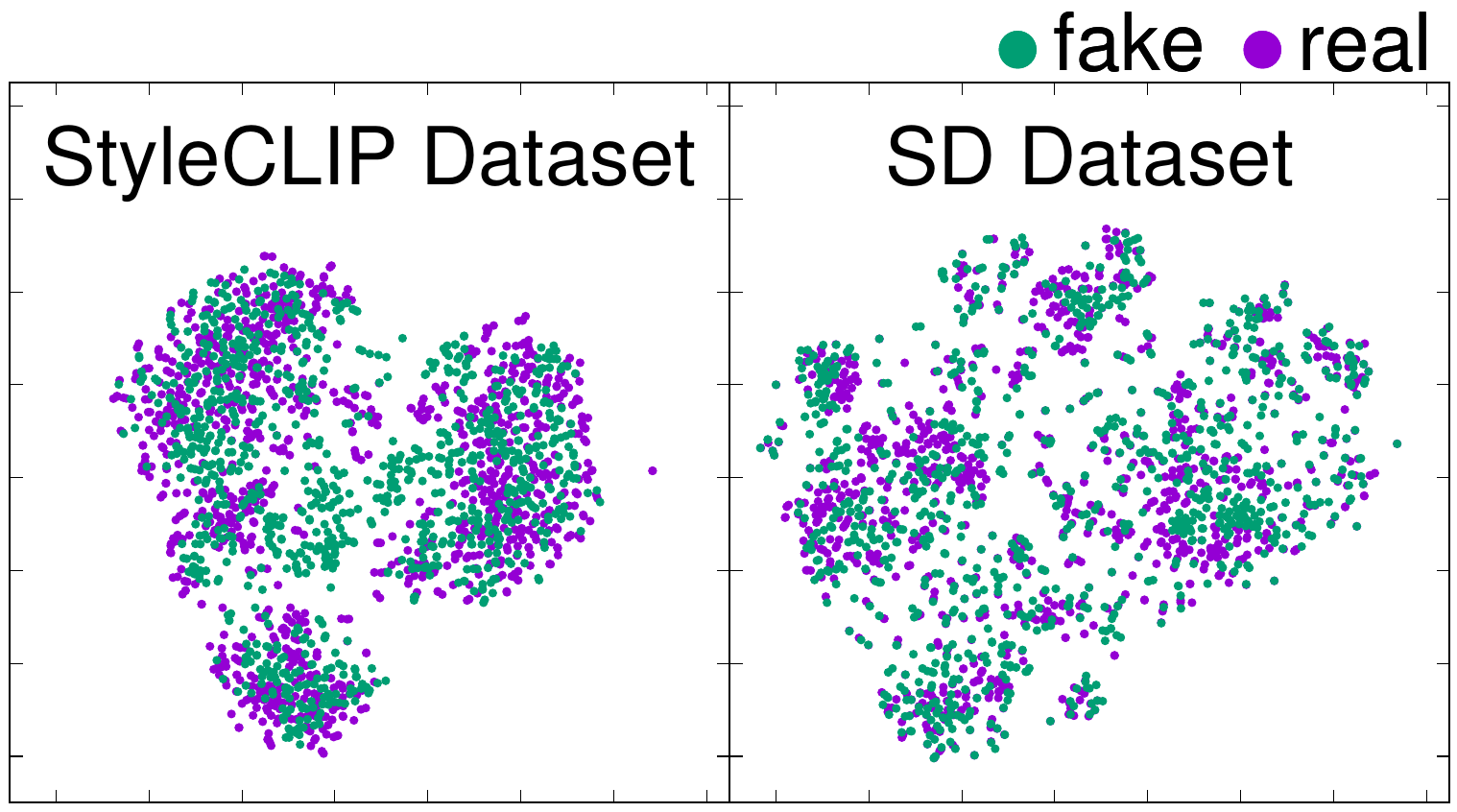}
\caption{Our dataset}
\end{subfigure}
\caption{TSNE plots of real and fake images used in \univclip{} defense (left) and our datasets (right). Fake and real images in the \uclip{} dataset are easier to separate as they are not controlled for content and quality.}
\vspace{-0.1in}
\label{fig:tsne_univclip}
\end{figure}

To understand the impact of \textit{not} controlling the content and quality of images, we take a closer look at the \uclip{} defense. Recall that \uclip{} uses the CLIP-ViT foundation model to extract features for detection.
Figure~\ref{fig:tsne_univclip} shows a TSNE visualization to capture how well \uclip{} can separate real and fake images using features extracted from CLIP-ViT. The plots on the left show the results on the (problematic) datasets used in the original work (\uclip{}), \ie fake images from the LDM~\cite{rombach2022high} and GLIDE~\cite{nichol2021glide} generative models, and real images from the LAION-400M~\cite{schuhmann2021laion} dataset. We find that LDM and GLIDE images have lower visual quality compared to our SD and StyleCLIP datasets.\footnote{The Kernel Inception Distance (KID) between real and fake images for the LDM-200 model and GLIDE used by \uclip{} is 0.023 and 0.017, respectively, while KID for our SD dataset is 0.008, \ie an order of magnitude lower. Lower KID indicates better fake image quality.} \uclip{} is able to almost perfectly separate these classes in this case. However, when we visualize the foundation model features on our SD and StyleCLIP datasets (plots on the right), we observe a drastically different result --- features from CLIP-ViT being unable to cleanly separate these classes. This suggests that only using foundation model features, without a linear classification layer, will not generalize well---which invalidates the fundamental claim made by \uclip{} that such features with a nearest neighbor search is sufficient for high generalization performance.

\textit{The high generalization performance of \uclip{} in the original work can be attributed to the lack of control of content and quality of the images}, which is likely overestimating its real-world performance---it is easier to separate fake images with heavy artifacts and different content distribution from real images.
However, for the rest of our evaluation, we fixed these problems that plague \uclip{}, retraining it on our SD and StyleCLIP datasets, where we control both quality and content of images. The trainable linear classification layer in our implementation helps to better discriminate between fake and real images.

\para{Limitation 2: Lack of adversarial evaluation.} An effective defense should be robust against an adaptive adversary. In a practical setting, this would be an attacker in a black-box setting (with no access to defense internals) who crafts ``adversarial'' fake images that evade detection (\ie classified as real). Such an attacker may also exploit some knowledge of the defense, \eg the defense uses frequency domain features or texture features. We find that existing work is severely lacking in this respect. \textit{\dct{} and \uclip{} do not conduct any evaluation in adversarial settings.} Among the remaining defenses, all except \patch{} and \defake{}, only conduct a basic robustness evaluation. This includes basic image manipulation schemes such as blurring, JPEG compression, downsampling, and noising/denoising. \patch{} studies an adaptive attack, but uses a white-box setting, where the attacker has full access to the defense internals. This is not the most practical setting because the defense internals may not be publicly released. Finally, \defake{} studies both white- and black-box attacks that show significant degradation in their detection performance.

\para{Limitation 3: Restricted image content types.} Prior work focused only on limited content types, \eg faces, animals, bedrooms, and buildings. Given the proliferation of text-to-image models, an attacker can generate fake images that capture \textit{any} type of content using a text prompt. We encourage the community to consider a broad range of content to study deepfake defenses. This can also present new technical challenges, as some defenses are designed only for faces~\cite{chai2020makes}. Newer datasets, such as LAION-5B~\cite{schuhmann2022laion}, have image-text pairs that include photographs, artistic paintings and illustrations covering a broad range of semantic content. Text captions from such datasets can be used to generate fake images that cover a wider range of content.

%% file: evaluation2.tex
\section{Defenses Against Evolving Threats}
\label{sec:evaluation}

\subsection{Evolving Threat 1: User-Customized Generative Models}
\label{sec:generalization_testing}

\noindent Our goal here is to understand the effectiveness of existing defenses in a threat landscape enabled by the democratization of generative AI technologies. The open source release of the Stable Diffusion model and the development of new low-cost generative model fine-tuning approaches have resulted in thousands of user-customized (\ie by Internet users) SD model variants on platforms like CivitAI and HuggingFace. Internet users are creating custom checkpoints of the base SD model for various reasons---to enhance image quality, realism, adapt to a new image dataset or to change the style of images.
We carefully choose representative user-customized models from this pool to evaluate the generalization performance of existing defenses on them. Note that we do not study the StyleCLIP model in this section due to its lack of widespread user-customization, which is a requirement for this analysis.

\begin{figure*} [t]
\centering
\includegraphics[scale=0.206]{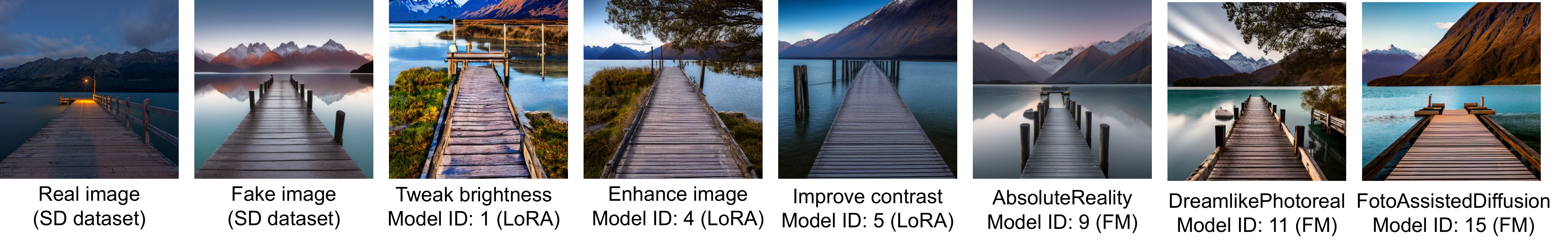}
\caption{Image samples for the caption ``Dawn at a jetty in Glenorchy, New Zealand." From left to right, first 2 images are real and fake images from our SD test set, the next 3 are LoRA images, followed by 3 FM images. Model IDs are explained in Table~\ref{tab:att_custom_lora} (Appendix~\ref{sec:user-created-model-details}). We can see content preservation with comparable quality across all samples.}
\label{fig:sd_variant}
\end{figure*}

\subsubsection{Collecting user-customized SD models} \label{sec:user_customized_sd_models} The traditional approach to create a custom model is to fine-tune the base model on a new dataset by updating all parameters (layers) of the model---known as \textit{Full Model fine-tuning (FM)}. However, this is computationally expensive and requires high-end GPU hardware to fine-tune SD on large datasets. Recently, a novel approach called \textit{Low-Rank Adaptation (LoRA)} was proposed to enable low-cost fine-tuning of generative models~\cite{hu2021lora}. For SD, LoRA is applied by adding a small number of trainable parameters to the cross-attention layers, which extracts correlations between images and text prompts. The rest of the SD model is kept frozen during fine-tuning. Using LoRA results in faster training time, requires less compute (can be run on consumer-level GPUs, \eg NVIDIA 2080TI), and the trained weights produce smaller files (order of MBs, compared to GBs for FM fine-tuning)~\cite{UsingLoR47:online}.

\textit{We only choose custom models that enhance the aesthetics of the images while preserving the semantic content and quality, compared to our SD dataset (training dataset for our defenses).}
It would be unfair to expect the defenses to generalize if the custom checkpoints entirely change the domain/content of the images seen during training. We choose a total of $16$ custom checkpoints, $8$ based on LoRA fine-tuning, and $8$ based on FM fine-tuning. Users fine-tuned all checkpoints from the SDv1.5 base model. We apply the LoRA weights using the Diffusers library~\cite{von-platen-etal-2022-diffusers} with $\alpha=0.5$, which is a scaling factor for the LoRA weights.\footnote{A lower scaling factor ensures that the semantic content is mostly preserved.} Table~\ref{tab:att_custom_lora} (Appendix~\ref{sec:user-created-model-details}) presents the details of each checkpoint. For example, there are checkpoints that enhance brightness, details, sharpness, contrast, reduces noise, and increases realism. Figure~\ref{fig:sd_variant} shows samples of images from both LoRA and FM checkpoints. More samples are in Figure~\ref{fig:sd_variant_extra} (Appendix). For each of the $16$ custom models, we generate $1K$ images using text prompts from our SD test dataset. These images serve as the testing set to measure generalization performance of existing defenses.

Before we proceed with evaluation of the defenses, we test whether the user-customized models indeed meet our requirements, i.e., preserving the semantic content and not degrading image quality. We use the following two metrics that have been shown to correlate with human judgement~\cite{kumari2023multi}:

(1) \textit{Semantic similarity:} We use the metric \textbf{CLIP-Score}~\cite{hessel2021clipscore}.
CLIP-Score measures the cosine similarity between an image and its text prompt using the CLIP-ViT-B/32 model~\cite{dosovitskiy2020vit}, \ie how well does an image capture its text prompt? The values range between 0 and 1, with higher values indicating better semantic similarity. We compute the CLIP-Score for each generated image using its associated text prompts (from our SD dataset). 
As a baseline for comparison, we calculate the CLIP-Score for fake images in our SD dataset. 
For the baseline images (from Realistic Vision), we obtain an Avg. CLIP-Score of $0.82$. The user-customized models have a similar CLIP-score value as well---Avg. CLIP-score values across LoRA and FM models are 0.81 and 0.81, respectively. This suggests that user-customized models preserve the semantics of the content.

(2) \textit{Image quality}: We use the \textbf{Kernel Inception Distance (KID)}~\cite{binkowski2018demystifying} metric to measure synthetic image quality.\footnote{Our dataset has 1k images each for fake and real. We use KID instead of FID~\cite{heusel2017gans}, because FID provides reliable estimates for only much larger datasets~\cite{karras2020training}.} KID measures the distribution distance between the real and fake image sets. The values are unbounded, and smaller values (closer to zero) indicate better synthetic image quality, \ie fake images match the distribution of the real images. For the baseline images from our SD dataset, we have a KID score of 0.008, and obtain an Avg. KID score of 0.006 and 0.008 for LoRA and FM models, respectively. The KID values are small (close to zero) and close to the baseline results, suggesting that these models are not degrading image quality.

\subsubsection{Defense generalization on user-customized SD models} \label{sec:user_custom_sd_results}
\begin{figure} [h]
\centering
\includegraphics[height=0.53\textwidth, width=0.48\textwidth]{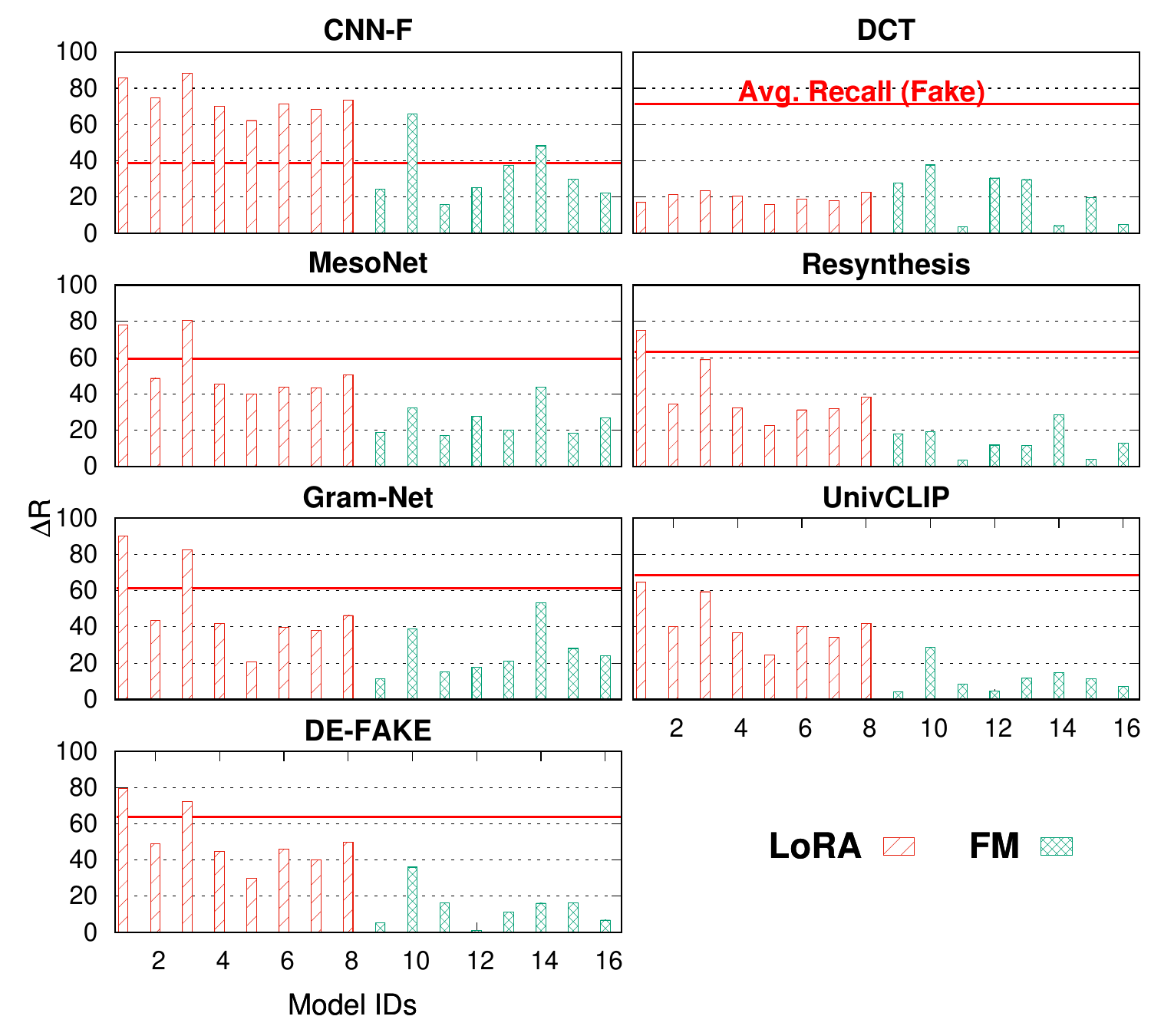}
\caption{Generalization performance of defenses measured using $\Delta R$. Model IDs 1 to 8 are LoRA models, and 9 to 16 are FM models. See Table~\ref{tab:att_custom_lora} (Appendix~\ref{sec:user-created-model-details}) for details. Red line in each plot shows the average Recall (fake class) over all 16 SD custom checkpoints.}  
\label{fig:lora_custom}
\end{figure}

We use the version of defenses trained on our SD dataset (Section~\ref{sec:defense-implementation}) for this evaluation. To measure defense performance, we compute $\Delta R$, calculated as the \textit{percentage degradation in Recall for the fake class}, compared to the baseline test performance of a defense.\footnote{We do not consider the Precision metric because no real images are required for this analysis.} In other words, $\Delta R = \frac{R1 - R2}{R1}$, where $R1$ is the Recall on our SD test set, and $R2$ is the Recall when applied to images from a user-customized SD model. The baseline Recall numbers ($R1$) are in Table~\ref{tab:before_after_defense_test}. \textit{A high $\Delta R$ would indicate that a defense performs poorly when applied to user-customized models.} Figure~\ref{fig:lora_custom} shows the results. The red line in each plot represents the average Recall across all user-customized model variants for the corresponding defense.

\textsc{\colorbox{lightblue}{Finding 1.}}
\textit{\textbf{All} defenses exhibit significant degradation in performance.} The average $\Delta R$ for defenses (across all the user-customized models) ranges from 19.69\% to 53.92\%. This performance degradation highlights the urgent need to develop new defenses or enhance existing defenses to improve generalization performance.

\textsc{\colorbox{lightblue}{Finding 2.}}
\textit{Solely relying on features from a foundation model is insufficient to achieve high generalization.} 
\univclip{} and \defake{} claim that using features from a foundation model (CLIP-ViT) is sufficient to achieve high generalization performance. However, our evaluation does not support this claim. Figure~\ref{fig:lora_custom} shows that both \uclip{} and \defake{} demonstrate significant performance degradation. For LoRA models, we observe an average $\Delta R$ of 42.66\% and 51.35\% for \uclip{} and \defake{}, respectively. For FM models, while the degradation is lower compared to LoRA, we see $\Delta R$ up to 28.87\% and 35.99\% for \uclip{} and \defake, respectively.
LoRA is a dominant low-cost fine-tuning strategy among Internet users, so this presents a real threat. Given our finding, one may wonder how \uclip{} demonstrated high generalization across diverse generative models in their work. We suspect that this can be attributed to their evaluation data sets that do not control the content and quality of the images (see analysis in Section~\ref{sec:defenselimitations}).

\textsc{\colorbox{lightblue}{Finding 3.}} \textit{Frequency domain features show the best generalization performance.} \dct{}, a defense based on the frequency domain features shows the most promise. It shows an average $\Delta R$ (across all models) of 19.69\%, which is the lowest $\Delta R$ among all defenses. This is because the frequency spectrum artifacts found in the LoRA and FM model images resemble those observed in images encountered during defense training (SD fake images), whereas real images do not exhibit such artifacts. We visualize the frequency spectrum of real and fake images with further explanation in Appendix~\ref{sec:freqartifact_sd}.

\textsc{\colorbox{lightblue}{Finding 4.}}
\textit{CNN-based defenses show the worst generalization performance.} Note that the \cnn{} and \meso{} and \texture{} defenses show high F1 scores ($>90\%)$ on our SD dataset. However, CNN-based defenses, even with specialized architectures, are unable to learn generalizable features. \cnn{}, \meso{} and \texture{}, all based on CNNs, exhibit the highest average $\Delta R$ (across all the user-customized models), ranging from 38.26\% to 53.92\%. These schemes claim to use data augmentation strategies to help with generalization. Nevertheless, data augmentation techniques are unable to compensate for features that do not generalize effectively.

\subsubsection{Improving generalization using content-agnostic features} 
\label{sec:content-agnostic}
Even though user-customized models preserve high-level image semantics, there may be distributional differences in low-level content. Ideally, defenses should focus on imperfections of the fake images, and not be derailed by changes in the content distribution. Based on this insight, we investigate whether enhancing an existing defense with ``content-agnostic'' features would improve generalization.

We enhance the \dct{} defense as it shows the most promise for generalization. Prior work has shown that the noise residual or the ``noise space'' of an image, \ie the residual image after removing all the content, can contain effective discriminatory patterns for deepfake detection~\cite{pu2020noisescope}.
\footnote{Many of the defenses we study (from Section~\ref{sec:defense_criteria}) claims to outperform schemes that rely only on noise residuals.} Inspired by this work, we use a SOTA denoising scheme, MM-BSN~\cite{zhang2023mm}, to extract noise residuals for each image in our SD dataset. From these noise residuals, we extract the log-scaled \dct{} features, which serve as our content-agnostic features. We then enhance our \dct{} scheme by concatenating existing \dct{} features (calculated over the entire content space) with our content-agnostic features and retrain the \dct{} scheme.

\textsc{\colorbox{lightblue}{Finding 5.}} \textit{Content-agnostic features can help boost generalization performance for deepfake detection.} The enhanced \dct{} scheme with content-agnostic features shows improvement, \ie there is a reduction in $\Delta R$ for 12 out of the 16 user-customized models. The average $\Delta R$ over all FM models reduces from 19.68\% to 14.29\%. For LoRA models, we see a lower improvement--average $\Delta R$ reduces from 19.69\% to 19.21\%. Content-agnostic features are promising. With better noise residual extraction schemes, combined with better learning schemes, generalization can be potentially further improved.

\subsubsection{Improving generalization using ensemble approaches} \label{sec:ensemble_with_dct} Since most defenses use different methods/architectures for their defenses, it is possible that defenses are learning different artifacts to separate real from fake. This can be leveraged using an \textit{ensemble approach} to improve generalization. We again build on the most effective \dct{} defense. We combine \dct{} with each of the remaining 6 defenses, and flag an image as fake if any one method predicts it to be fake. Of course, such an ensemble scheme can degrade Precision, \ie increase false positives. Therefore, for this analysis, we report Precision, Recall and F1 score metrics for the fake class.

\textsc{\colorbox{lightblue}{Finding 6.}} \textit{Combining domain-specific features (\ie features known to identify imperfections in fake images) with features from a foundation model improves generalization.} Table~\ref{tab:dct_ensemble} shows the Average Precision, Recall and F1 scores (for the fake class) over all the 16 user-customized models for each defense variation. The top row shows the performance of the \dct{} scheme without an ensemble approach. 
Combining \dct{} with \uclip{} shows the largest improvement in F1 score---up to 86.25\%. Recall that only using foundation model features does not improve generalization performance (Finding 2). However, foundation model features in conjunction with domain-specific features (\ie frequency features) can improve generalization performance. Also note that it is easier to build an ensemble defense when we are extracting features from a pretrained model (\ie foundation model), compared to a defense like \texture{} (which also comes close in performance). 
The \defake{} defense which also uses features from a foundation model, performs similar to \dct{}+\uclip{}. We also explored combining the second-best defense (Figure~\ref{fig:lora_custom}), \uclip{} with other defenses (other than \dct{}) to find better ensembles. We did not see any other combination that achieves an F1 score higher than \dct{}+\uclip{}, which confirms the supremacy of combining both types of defenses. Detailed results are in Appendix~\ref{sec:univclip_ensemble}.

\begin{table}[t]
\centering
\normalsize
\setlength{\tabcolsep}{3pt}
\setlength\extrarowheight{3pt}
\begin{tabular}{l"c|c|c}
\textbf{Defenses} & \textbf{Precision (\%)} & \textbf{Recall (\%)} & \textbf{F1 (\%)} \\ \hlinewd{1.1pt}
\dct{} & 84.74 & 71.32 & 77.27 \\ 
\dct{} + \texture{} & 86.63 & 86.01 & 86.21 \\ 
\dct{} + \resyn{} & 78 & 87.13 & 82.26 \\ 
\dct{} + \cnn{} & 85.64 & 77.64 & 81.29 \\ 
\dct{} + \meso{} & 83.70 & 85.63 & 84.57 \\ 
\dct{} + \uclip{} & 82.97 & 89.95 & 86.25 \\ 
\dct{} + \defake{} & 83.26 & 89.46 & 86.16
\end{tabular}%
\caption{Improving generalization performance by creating an ensemble model by combining \dct{} defense with the other 6 defenses. We show the average scores across the 16 SD custom checkpoints.}
\label{tab:dct_ensemble}
\end{table}

%% file: evaluation3.tex
\subsection{Evolving Threat 2: Adversaries Leveraging Vision Foundation Models}
\label{sec:advattack}

\subsubsection{Using foundation models to craft adversarial fake images}
\label{sec:advattack-methodology}
We propose a simple black-box attack to craft ``adversarial'' fake images by leveraging vision foundation models. We assume that the attacker has already created a fake image that is deceptive or damaging, but it can be caught by a deepfake detector. To bypass detection or to create an adversarial version of this fake image, the attacker chooses a semantic property of this image for adversarial manipulation that still preserves their goal of deception. 
For example, a face image created to build a fake social media profile is misclassified as real because the facial expression has undergone adversarial manipulation. Here, the main requirement for the attack may be to have a realistic looking profile picture, which is still achieved, despite the manipulation.
The semantic property to be manipulated is expressed using a \emph{\textbf{text prompt}}. Note that our approach does not add any adversarial noise and only adversarially manipulates the content to match the content described by the text prompt, \eg a prompt that says ``a smiling face'' should craft an adversarial fake image with a smiling face.

Our idea is to adversarially update the weights of a fake image generator, guided by a surrogate deepfake classifier that is implemented using a foundation model. Such an adversarially trained generator can produce adversarial fake images. We use foundation models to build our surrogate classifier for the following reasons: (1) Previous work shows that foundation model features are effective for deepfake detection, \eg \uclip{} and \defake{} (also see discussion in Section~\ref{sec:background}). (2) Foundation models being pretrained models, provide a ready-to-use model to easily build a deepfake classifier. (3) Widespread availability of diverse foundation models, provides several options for the attacker to instantiate different types of surrogate deepfake classifiers.

We demonstrate our attack using the StyleCLIP generative model (see Section~\ref{sec:only_gen_models}) on face content. StyleCLIP uses the StyleGAN2 image generator, and we focus on face images. \textit{Note that this is a generic attack and does not use any specific properties of the StyleCLIP (or StyleGAN2) model}.
 For this attack, we use 3 foundation models that are image encoders: (1) EfficientNet~\cite{tan2019efficientnet}: 5.3M parameters, trained on 14M images, (2) ViT~\cite{dosovitskiy2020vit}: 86M parameters, trained on 14M images, (3) CLIP-ResNet~\cite{radford2021learning}: 623M parameters, trained on 400M images. Note that EfficientNet is one of the smallest foundation models, and can even run on mobile devices.

We consider an image generator $G(\theta)$ that takes as input an existing fake image $x$ and a text prompt $p$ (to guide generation)  to generate a new image $x'=G(x,p;\theta)$. The surrogate deepfake classifier is $M$ with an associated likelihood probability function $p_M$. The attacker uses the surrogate model to adversarially update the weights of $G(\theta)$ to create $G(\theta_{adv})$, such that $G(x,p;\theta_{adv})=x_{adv}$, where $x_{adv}$ is the adversarial fake image. The surrogate model $M$ is frozen. The generator $G$ is adversarially trained to minimize the following loss objective $L$:
\begin{equation}
    L = \gamma * L_{cls} + \delta * L_{percept}
\end{equation}
where $L_{cls}$ is the classification loss, $L_{percept}$ is a perceptual loss term, and $\gamma$ and $\delta$ are associated coefficients. 

\begin{equation}
L_{cls} = \mathbb{E}_{p_{data(x')}}[l(p_{M}(c|x'),c)]
\end{equation}
$L_{cls}$ is implemented as cross-entropy loss $l(.)$ using the surrogate classifier $M$. $c$ is the ground-truth label which is set to the \textit{real} class so that the generated image is classified as \textit{real}. The perceptual loss $L_{percept}$ is used to ensure that the image quality is not degraded. We use the intermediate representations extracted from an ImageNet-trained VGG network~\cite{simonyan2014very} to calculate the perceptual loss~\cite{zhang2018unreasonable} using $x$ and $x'$. Note that our threat model does not require the facial identity to be preserved and therefore does not include any loss objectives to preserve identity. Based on $L$, the generator is trained for 50 iterations per image $x$.

We use the $1K$ fake images from our StyleCLIP test set (Section~\ref{sec:defense-implementation}) to create an adversarial version of each image using the above methodology. For each image, we choose a randomly chosen target prompt from a pool of 8 prompts, \eg \textit{a smiling face}, \textit{a face with lipstick}, \textit{a face with glasses}, etc. (see Appendix~\ref{sec:adv-attack-details}). To build surrogate classifiers using the chosen foundation models (EfficientNet, ViT, CLIP-ResNet) we follow a simple methodology similar to \uclip{} with some minor variations. To train the surrogates, the attacker can use fake images from their current generator $G$, and any publicly available real image datasets that capture the desired content. We used a subset of real images from the FFHQ dataset~\cite{karras2019style} with no overlap with the training set of the defenses. All 3 surrogate classifiers achieve high test F1 scores ranging from 98.15\% to 98.38\%. Surrogate models are trained only once with images from $G$ and further frozen, while we adversarially update $G$. Details of the training configurations of the surrogates and the generator are in Appendix~\ref{sec:adv-attack-details}.

\begin{table*}[!t]
\centering
\small
\setlength{\tabcolsep}{1pt}
\setlength\extrarowheight{3pt}
\begin{tabular}{l"cccccccc"cc}
\multirow{2}{*}{\textbf{\begin{tabular}[c]{@{}l@{}}Surrogate\\ model\end{tabular}}} & \multicolumn{8}{c"}{\textbf{$\Delta R$ for fake in \%}} & \multicolumn{2}{c}{\textbf{Semantic and quality metrics}} \\ \cline{2-11} 
 & \multicolumn{1}{c|}{\textbf{\texture{}}} & \multicolumn{1}{c|}{\textbf{\dct{}}} & \multicolumn{1}{c|}{\textbf{\resyn{}}} & \multicolumn{1}{c|}{\textbf{\cnn{}}} & \multicolumn{1}{c|}{\textbf{\meso{}}} & \multicolumn{1}{c|}{\textbf{\uclip{}}} & \multicolumn{1}{c|}{\textbf{\patch{}}} & \textbf{\defake{}} & \multicolumn{1}{c|}{\textbf{\clsr{}}} & \textbf{KID\textsubscript{Fake}} \\ \hlinewd{1.1pt}
\textbf{EfficientNet} & \multicolumn{1}{c|}{41.67} & \multicolumn{1}{c|}{57.43} & \multicolumn{1}{c|}{44.58} & \multicolumn{1}{c|}{40.06} & \multicolumn{1}{c|}{44.82} & \multicolumn{1}{c|}{4.56} & \multicolumn{1}{c|}{28.37} & 75.76 & \multicolumn{1}{c|}{0.675} & 0.007 \\ 
\textbf{ViT} & \multicolumn{1}{c|}{36.04} & \multicolumn{1}{c|}{53.41} & \multicolumn{1}{c|}{37.39} & \multicolumn{1}{c|}{32.23} & \multicolumn{1}{c|}{34.67} & \multicolumn{1}{c|}{4.66} & \multicolumn{1}{c|}{13.47} & 78.43 & \multicolumn{1}{c|}{0.675} & 0.005 \\ 
\textbf{CLIP-ResNet} & \multicolumn{1}{c|}{47.20} & \multicolumn{1}{c|}{88.35} & \multicolumn{1}{c|}{73.96} & \multicolumn{1}{c|}{70.85} & \multicolumn{1}{c|}{76.08} & \multicolumn{1}{c|}{12.47} & \multicolumn{1}{c|}{40.96} & 80.04 & \multicolumn{1}{c|}{0.671} & 0.017
\end{tabular}%
\caption{Evaluation of the defenses on the adversarial fake images created using our attack. We report $\Delta R$ for fake class, and for quality metrics, we report \clsr{} and KID\textsubscript{Fake}.}
\label{tab:adversarialattackresults}
\end{table*}

\begin{figure*}[t]
\centering
\begin{subfigure}[t]{2.0\columnwidth}
\includegraphics[height=0.11\textwidth,width=\columnwidth]{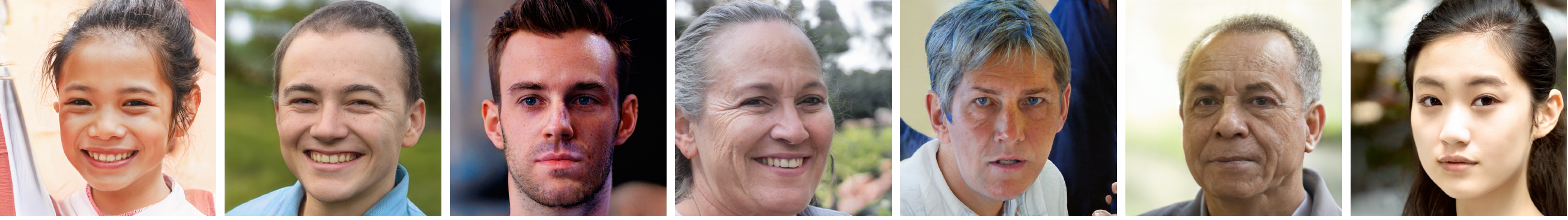}
\caption{Source fake}
\end{subfigure}
\\ 
\begin{subfigure}[t]{2.0\columnwidth}
\includegraphics[height=0.11\textwidth,width=\columnwidth]{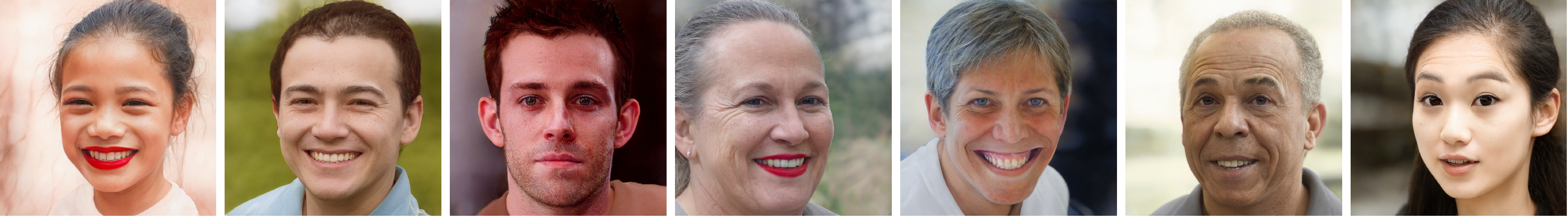}
\caption{Non-adversarial fake}
\end{subfigure}
\begin{subfigure}[t]{2.0\columnwidth}
\includegraphics[height=0.11\textwidth,width=\columnwidth]{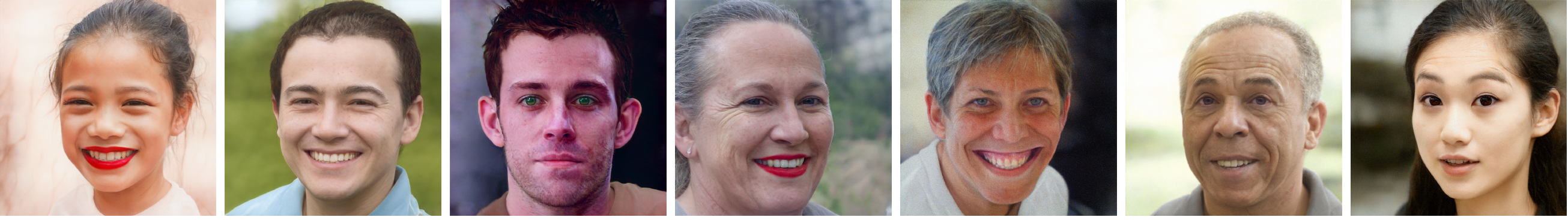}
\caption{Adversarial fake}
\end{subfigure}
\caption{Images in top row are the source fake images from our test set that are fed to the generator along with the prompt. Middle row shows the corresponding non-adversarial fake images, and the bottom row shows adversarial fake images. From left to right in row (c), the first 2 images are from EfficientNet, next 2 are from ViT, and the rightmost 3 images are from CLIP-ResNet surrogates. We explain more in Appendix~\ref{sec:adv-attack-details}.}
\label{fig:quality_adv_attack}
\end{figure*}

\subsubsection{Computational cost of our attack} On average for all 3 surrogate models, it takes 39 seconds to generate an adversarial image using an NVIDIA A100 GPU. The cost of using such a GPU for 1 hour is $\$1.1$.\footnote{\href{https://lambdalabs.com/service/gpu-cloud\#pricing}{https://lambdalabs.com/service/gpu-cloud\#pricing}} 
\textit{This enables generation of around 840 images with just $\$10$.} Therefore, foundation models enable a viable practical approach to craft adversarial fake images.

\subsubsection{Attack effectiveness} 
\label{sec:adv-attack-effectiveness}
We consider a challenging attack setting where all 8 defenses (Section~\ref{sec:defense_criteria}) are optimized (trained) to detect images from the attacker's generator. In other words, we use the version of the defenses trained on our StyleCLIP dataset, and the attacker adversarially updates the StyleCLIP generator.

Attack success is measured using 3 metrics: \\
(1) \textit{Percentage degradation in Recall $\Delta R$}: This is similar to the $\Delta R$ metric in Section~\ref{sec:generalization_testing}. $\Delta R=\frac{R1-R2}{R1}$, where $R1$ is the Recall on our StyleCLIP test set (Table~\ref{tab:before_after_defense_test}), and $R2$ is the Recall when adversarial fake images are used. A high value of $\Delta R$ indicates high attack success, \ie more degradation in defense performance. \\
(2) \textit{Measuring semantic similarity using CLIP-Score:} We use the same CLIP-Score metric from Section~\ref{sec:generalization_testing} to measure how well an adversarial fake image matches the desired content expressed using the prompt. A successful adversarial fake image should have a CLIP-Score that is similar to the fake image produced using the same prompt without the attack, \ie a non-adversarial fake image. \\
(3) \textit{Measuring fake image quality using KID:} KID (also used in Section~\ref{sec:generalization_testing}) can no longer be calculated between adversarial fake images and real images, because the content has been explicitly manipulated using a text prompt. Instead, we calculate KID\textsubscript{Fake} between our set of adversarial fake images and fake images produced using the same set of prompts, but without performing an adversarial attack. Smaller values of KID\textsubscript{Fake} (close to zero) would indicate higher image quality. In other words, smaller values of KID would indicate that adversarial fake images are similar in quality to the fake images produced without adversarially updating the generator.

Table~\ref{tab:adversarialattackresults} presents the attack results.

\textsc{\colorbox{lightblue}{Finding 7.}} \textit{Adversarial attacks using a foundation model can significantly degrade the performance of all defenses.} From Table~\ref{tab:adversarialattackresults} we see that all defenses exhibit a degradation in performance for all surrogate models, with $\Delta R$ being the highest when using the largest foundation model, CLIP-ResNet. We also see that 5 out of 8 defenses exhibit a $\Delta R$ higher than 70\%. The KID\textsubscript{Fake} values are small and close to zero, suggesting no significant degradation in image quality~\cite{binkowski2018demystifying} for attacks.
All three attacks result in adversarial fake images with CLIP-Scores in the range 0.671-0.675, which is similar to the CLIP-Score for non-adversarial fake images of 0.672 (not shown in Table~\ref{tab:adversarialattackresults}). 

It is evident from our findings that an attacker can easily exploit a publicly accessible foundation model to evade deepfake defenses \textit{that employ a variety of techniques.} Figure~\ref{fig:quality_adv_attack} shows several examples of the source fake images (\ie the one fed to the generator with the prompt), and the corresponding non-adversarial and adversarial fake images. We can see that the source images have noticeable content differences from the other sets, due to the prompt translating the source image. We further discuss the subtle content changes between the adversarial and non-adversarial fake images in Appendix~\ref{sec:adv-attack-details}.

\begin{figure} [!t]
\centering
\includegraphics[width=0.49\textwidth]{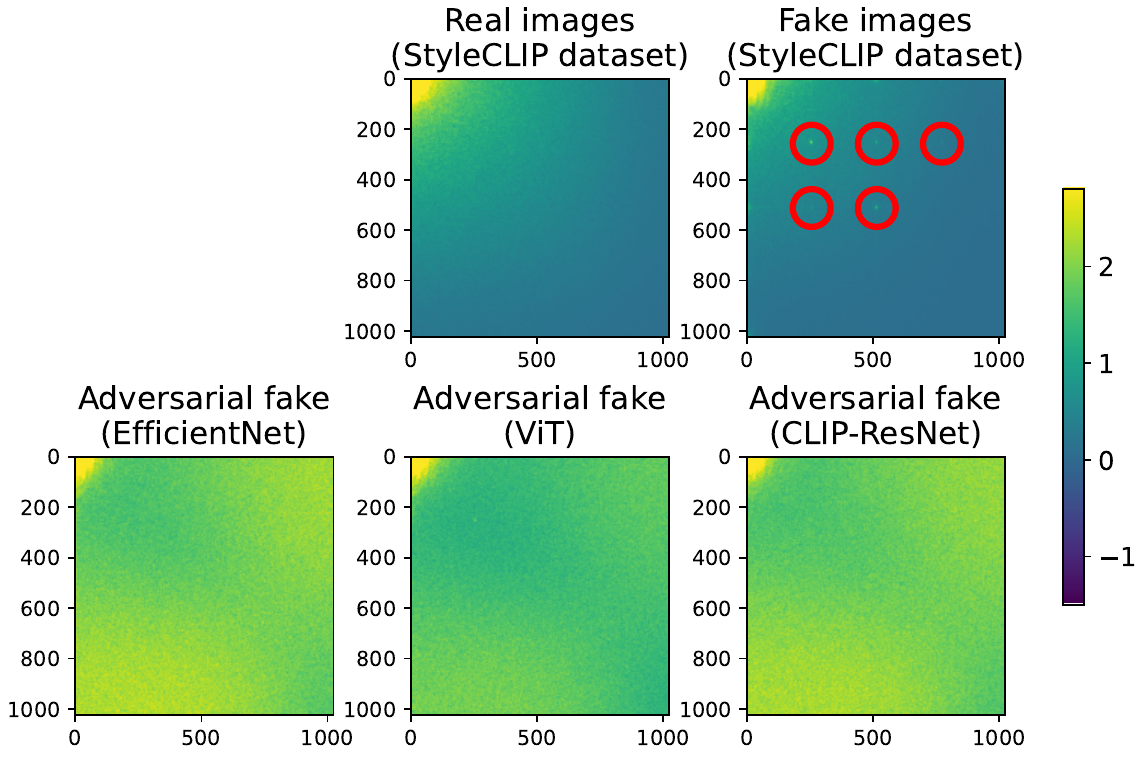}
\caption{Average log(DCT) spectrum calculated over 1$K$ images for each category. Top row shows frequency spectrum for real and fake images from our StyleCLIP dataset, and the bottom row shows adversarial fake images created using our 3 surrogate models. Red circles in ``Fake images (StyleCLIP dataset)" highlight the artifacts. }
\label{fig:dct_freq_graph}
\end{figure}

\textsc{\colorbox{lightblue}{Finding 8.}} \textit{Defense based on frequency features is the weakest against adversarial attacks using foundation models}. Although we see \dct{} as the strongest defense for generalization in Sec~\ref{sec:generalization_testing}, it is the weakest against adversarial attacks, exhibiting a high $\Delta R$ of up to 88.35\%. 
To understand this result, we visualize the frequency spectrum of real and fake images in Figure~\ref{fig:dct_freq_graph}. We can see that adversarial fake images mimic the frequency spectrum patterns of real images showing no clear artifacts, unlike the (source) fake images. It is fascinating that this was achieved without using a surrogate model based on frequency features. Our results further highlight the shortcomings of the frequency domain in an adversarial setting and the strength of a foundation model to thwart a variety of defenses. 

\textsc{\colorbox{lightblue}{Finding 9.}} \textit{Defense using foundation model features shows the most resilience}. \uclip{} shows the highest resilience against attacks across all 3 surrogate models. \uclip{} exhibits a $\Delta R$ less than 5\% against the two smaller foundation models (EfficientNet and ViT), but starts to decline in performance when the CLIP-ResNet foundation model is used as the surrogate. This leads to our next analysis to investigate whether defenses using foundation models trained on a larger dataset can provide better resilience. Note that \defake{} shows significant degradation against all surrogate models. While both \uclip{} and \defake{} use image features from a foundation model, \defake{} additionally uses text features (by automatically captioning the fake image) from a foundation model. We suspect that the poor performance of \defake{} can be attributed to the use of text features that may not be robust. Hence, using only image features from foundation models (similar to \uclip{}) is a better strategy.

\subsubsection{Improving adversarial robustness of defenses using foundation models trained on larger datasets}
\label{sec:arms-race}
Inspired by the \uclip{} defense and Finding 9, we create another defense similar to \uclip{}, but using a foundation model trained on a dataset larger than that used by CLIP-ViT (foundation model used by \uclip{}). We train a new defense called \textbf{\textit{\ularge{}}} which is trained using features from the foundation model OpenCLIP-ConvNext-Large~\cite{laionCLI71:online}. This foundation model is pretrained on 2B image-text pairs~\cite{schuhmann2022laion} and has 351M parameters. The training methodology for this new defense is similar to that of \uclip{}. We trained \ularge{} on our StyleCLIP dataset (Section~\ref{sec:defense-implementation}) for 30 epochs with the Adam optimizer using a learning rate of $1e-3$. This defense achieves a high test set Precision, Recall, and F1 score of 97.9\%, 97.6\% and 97.75\%, respectively.

\textsc{\colorbox{lightblue}{Finding 10.}} \textit{Defenses using more powerful foundation models can achieve better adversarial resilience.}
Table~\ref{tab:customdefeval} shows the results of adversarial attacks against our defense \ularge{}. There is almost no performance degradation, with $\Delta R$ less than 0.41\% for the 3 surrogate models. This result raises a new question: What if the adversary leverages a more effective foundation model (\ie pretrained on a larger dataset), compared to the defender, to build the surrogate? If a surrogate with a more effective foundation model leads to a higher attack success, then this can result in an arms race depending on who uses a better foundation model (among attacker and defender).\footnote{We are unable to implement an adversarial attack that uses the OpenCLIP-ConvNext-Large foundation model as the surrogate because of incompatibility in PyTorch versions used by StyleCLIP and OpenCLIP.}

\begin{table}[!t]
\centering
\normalsize
\setlength{\tabcolsep}{3pt}
\setlength\extrarowheight{2pt}
\begin{tabular}{l"c}
\textbf{\begin{tabular}[c]{@{}l@{}}Surrogate\\ model\end{tabular}} & \textbf{\begin{tabular}[c]{@{}c@{}}$\Delta R$ for fake (\%)\\for \ularge{} defense\end{tabular}} \\ \hlinewd{1.1pt}
\textbf{EfficientNet} & 0.10 \\ 
\textbf{ViT} & 0.41 \\ 
\textbf{CLIP-ResNet} & 0.10
\end{tabular}%
\caption{Defense performance of the \ularge{} defense measured using $\Delta R$ against attacks using 3 different surrogate models.}
\label{tab:customdefeval}
\end{table}

\begin{table*}[!t]
\centering
\normalsize
\setlength{\tabcolsep}{3pt}
\setlength\extrarowheight{3pt}
\begin{tabular}{l"ccc"cc}
\multirow{2}{*}{\textbf{\begin{tabular}[c]{@{}l@{}}Surrogate\\ model\end{tabular}}} & \multicolumn{3}{c"}{\textbf{$\Delta R$ (before/after) adversarial training in \%}} & \multicolumn{2}{c}{\textbf{\begin{tabular}[c]{@{}c@{}}Semantic \&\\ quality metrics\end{tabular}}} \\ \cline{2-6} 
 & \multicolumn{1}{c|}{\textbf{\texture{}}} & \multicolumn{1}{c|}{\textbf{\uclip{}}} & \textbf{\patch{}} & \multicolumn{1}{c|}{\textbf{\clsr{}}} & \textbf{KID\textsubscript{Fake}} \\ \hlinewd{1.1pt}
\textbf{\begin{tabular}[c]{@{}l@{}}(Adv. trained)\\ CLIP-ResNet\end{tabular}} & \multicolumn{1}{c|}{47.20 / 3.43} & \multicolumn{1}{c|}{12.47 / 1.12} & 40.96 / 15.57 & \multicolumn{1}{c|}{0.669} & 0.01
\end{tabular}%
\caption{Defense performance in $\Delta R$ for the top-3 defenses from Table~\ref{tab:adversarialattackresults}, and quality measured using CLIP-Score and KID\textsubscript{Fake}. $\Delta R$ is shown for defenses before and after adversarial training. The attack uses an adversarially trained surrogate.}
\label{tab:evaladvtrain}
\vspace{-1ex}
\end{table*}

\subsubsection{Improving adversarial robustness of defenses using adversarial training}
\label{sec:adv_training}
A popular strategy to build resilience against adversarial attacks is to perform \textit{adversarial training} of the defense classifier~\cite{goodfellow2014explaining}. In this case, the defense classifier is trained/fine-tuned on adversarial samples generated by the attacker. The assumption is that the defender can collect such adversarial samples after deployment. However, adversarial training is known to be vulnerable to further adaptations by the attacker~\cite{zhang2019limitations}. We study such a dynamic setting in this section---the defender adversarially trains their deepfake classifier, but subsequently, the attacker also adapts and uses an adversarially trained surrogate to craft a new distribution of adversarial samples. One would expect that such an adaptive attack can still significantly degrade the performance of the adversarially trained defense (as the attacker has also adapted).

For the attack, we use the most effective surrogate from Section~\ref{sec:adv-attack-effectiveness}---CLIP-ResNet. For defenses, we use \texture{}, \univclip{}, and \patch{}, which showed comparatively smaller degradation in performance, compared to the other defenses (Table~\ref{tab:adversarialattackresults}).

To adversarially train the current version of our defenses and the surrogate (used for attack) classifiers, we fine-tune them on a new \textit{adversarial StyleCLIP dataset} which is a balanced dataset of $5K$, $2K$, and $2K$ images for training, validation, and testing, respectively. The adversarial fake images are generated using the CLIP-ResNet surrogate. We use the same adversarial fake images to train both the defense and the surrogate because in practice one would expect the defender to have access to the adversarial samples for adversarial training. We ensure that the attacker and the defender use a disjoint set of images for the real class (all drawn from the FFHQ dataset). After adversarial training, both the surrogate classifier and the 3 defense classifiers achieve high test set F1 scores on the adversarial StyleCLIP dataset, ranging from 94.58\% to 99.40\%.

Next, we craft a new set of adversarial fake images (using the adversarial trained surrogate) to test the resilience of the defenses (with adversarial training). Note that, the attack is adaptive because of retraining the surrogate on adversarial images. We generate $1K$ adversarial images using a similar methodology as before (Section~\ref{sec:advattack-methodology}). Table~\ref{tab:evaladvtrain} shows the performance of the 3 defenses on this new distribution of adversarial fake images.

\textsc{\colorbox{lightblue}{Finding 11.}} \textit{Adversarial training can be an immediate strategy to improve adversarial resilience against our attack.} From Table~\ref{tab:evaladvtrain}, we see that $\Delta R$ has dropped significantly for all 3 defenses after adversarial training, despite being set up against an adversarially trained attacker.  For example, \texture{} improves from $\Delta R$ of 47.20\% to 3.43\%, while \uclip{} is now nearly fully resilient to adversarial attacks with a $\Delta R$ of only 1.12\%.
However, these results should not be taken as a message that adversarial training is a robust measure against such attacks. First, defenses still exhibit some performance degradation. Second, there may be alternative adaptive strategies by the attacker that substantially alters the distribution of the adversarial samples---such examples can potentially still disrupt these defenses.

%% file: related-work.tex
\section{Related Work}
\label{sec:discussion}

\noindent We already discuss defenses in Section~\ref{sec:defense_criteria}. Here, we focus on related work covering \emph{attacks on deepfake defenses.}

Some of the existing works~\cite{hou2023evading, carlini2020evading} 
to evade deepfake detection focus on adding adversarial noise to the images in the pixel space in both white- and black-box scenarios. Such attacks tend to add visible adversarial noise that degrades image quality~\cite{jia2022exploring, li2021exploring}. Our method does not add any adversarial noise.

Other attacks rely on eliminating specific artifacts from the fake images through post-processing to evade detection. For example, checkerboard artifacts in images produced by GAN models~\cite{Schwarz2021OnTF} can be countered by post-processing attack pipelines~\cite{wesselkamp2022misleading, huang2020fakepolisher} that remove such artifacts. Targeted removal of specific artifacts does not guarantee evasion against all defenses. We study defenses that use different methods targeting different sets of artifacts in fake images.

Carlini et al.~\cite{carlini2020evading} conducted a preliminary evaluation of an adversarial attack strategy that does not require the addition of adversarial noise to the images. Instead, adversarial perturbations are applied to the latent space of a StyleGAN generator. However, this attack assumes a white-box setting, unlike our attack, which considers a black-box setting. Li et al.~\cite{li2021exploring} and Jia et al.~\cite{jia2022exploring} also study adversarial attacks without adding adversarial noise in both white-box and black-box settings. However, these methods either target a specific architectural feature of the image generator or focus on adding adversarial perturbations to specific feature spaces of the images, \eg in the frequency domain. Our attack is more generic and does not use any specific properties of the generator or target specific feature spaces, thereby enabling broader applicability.

None of the above works systematically study the impact of using different foundation models to create surrogate classifiers. More importantly, these attacks are not thoroughly tested against a variety of state-of-the-art defenses.

%% file: conclusion.tex
\section{Future Work}
\label{sec:future-work}

\noindent We discuss several new directions for future work.

\textit{(1) New directions to improve generalization performance against user-customized image generators.} We find that combining image features from foundation models with domain-specific features, \eg frequency-based features, can significantly enhance the ability to generalize to user-customized deepfake generators. Our work also highlights the potential of using content-agnostic features that can capture imperfections in noise residuals for improved generalization. Future work can explore more effective methods to leverage these features for improved generalization.

\textit{(2) Building robustness against adversarial attacks powered by foundation models.} Our new attack highlights the ease with which attackers can leverage foundation models to fool SOTA detectors. We encourage the community to explore the following directions to address this pressing challenge: (a) Further explore how defenders can build and leverage highly effective foundation models to tilt the arms race in favor of the defender. We find that a defender using a more powerful foundation model, \ie pretrained on a larger dataset (compared to the attacker), shows improved adversarial resilience. (b) Explore novel adversarial training strategies to enhance adversarial resilience. Our analysis shows significant potential in adversarial training strategies even against an adaptive attacker.

\textit{(3) Generative model customization techniques continue to evolve, thus further expanding the deepfake threat surface.}
We only studied the threat of users customizing generative models using LoRA and FM fine-tuning strategies. This space is rapidly evolving with several Parameter Efficient Fine-tuning (PEFT)~\cite{PEFT20:online} methods being developed, \eg{} DreamBooth~\cite{ruiz2023dreambooth} and ControlNet~\cite{zhang2023adding}. Users can also combine multiple LoRA model weights to create a single custom model~\cite{EggFusio74:online}. Understanding how defenses generalize to these other customization strategies can be further explored.

\textit{(4) We need deepfake datasets covering a wide variety of content types to train and evaluate deepfake defenses.} Section~\ref{sec:defenselimitations} highlights the limitations of existing work that focuses only on certain content types. Our community can create new deepfake datasets based on datasets such as LAION-5B~\cite{schuhmann2022laion} which contains 5B image-text pairs, where the text captions can be used to generate new fake image datasets. We created such a dataset, but it is limited in size (our SD dataset in Section~\ref{sec:defense-implementation}). Creation of new large datasets can also accelerate the development of new \textit{content-agnostic defenses} (Section~\ref{sec:content-agnostic}).

\textit{(5) Foundation model-powered adversarial attacks can be more sophisticated and effective.} Foundation models will continue to evolve rapidly, learning more effective patterns from Internet-scale data. The evolution of these models alone can potentially further improve the performance of our attack without requiring any changes to our attack methodology. Future work can also explore more advanced attack strategies using foundation models that exploit specific properties of the image generator.
We only demonstrated our attack using the StyleCLIP generator. Our attack can also be applied to the Stable Diffusion model using classifier-guided image generation techniques~\cite{kim2022refining}.

\textit{(6) Our simple attack method using foundation models can be used to benchmark adversarial robustness of new defenses.} We highlight the lack of proper adversarial evaluations in existing defense studies (Section~\ref{sec:defenselimitations}). This is likely due to the community lacking established methods to adversarially probe a deepfake defense. We present a simple method using foundation models that can be easily applied to new defenses.

\section{Conclusion}
\label{sec:conclude}
\noindent Deepfake images continue to pose a serious threat, for which several highly effective machine learning-based defenses have been proposed. In this work, we showed that these advances in defenses face a serious challenge due to two recent developments in machine learning---emergence of low-cost generator customization schemes which enable attackers to create a large variety of deepfake generators, and the emergence of vision foundation models which can be integrated with existing generators to craft adversarial fake images. 
Using 16 user-customized SD models, we show that existing defenses struggle to maintain their high accuracy in detecting deepfakes. We also identify their vulnerabilities against our adversarial attack, 
 where we only make meaningful content manipulations without adding adversarial noise to the image. One of the main insights of our work is understanding the consequence of foundation models and their continuous advances on deepfake detection. We encourage further research on better leveraging foundation models for deepfake detection based on our findings.

\section*{Ethics Statement}
\noindent We only used data and pretrained models that have been publicly shared for research purposes. We did not use human subjects in our research. All attacks were conducted in controlled lab settings, and no deployed services were affected using our fake images. We believe that the benefits of our work far outweigh any potential harm caused by the knowledge of our attacks.

\section*{Acknowledgements}

\noindent This work was supported in part by the NSF under awards 1943351, 2231002 and the Commonwealth Cyber Initiative, an investment in
the advancement of cyber R\&D, innovation, and workforce development. The views and findings solely reflect those of the authors and should not be seen as representative opinions of any funding agency, given the absence of any financial conflicts.

%% file: appendix.tex


\begin{appendices}
\label{sec:appendix}

\begin{figure*}[h!]
\centering
\includegraphics[scale=0.2]{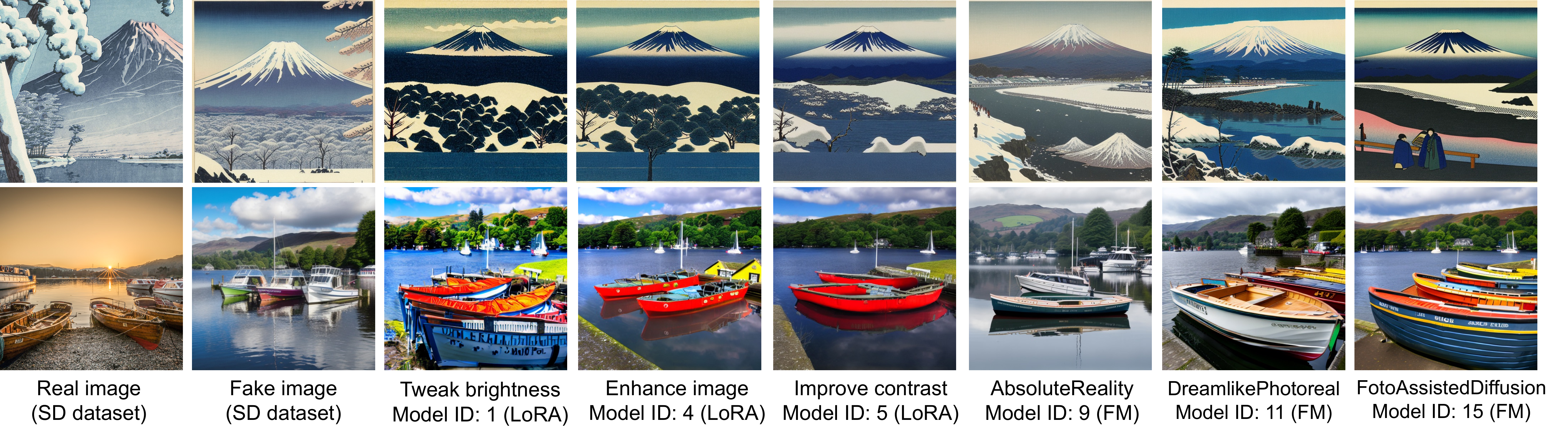}
\caption{From left to right: first 2 images are real and fake images from our SD test set, the next 3 are LoRA images, followed by 3 FM images. Model IDs are explained in Table~\ref{tab:att_custom_lora} (Appendix~\ref{sec:user-created-model-details}). We can see content preservation with comparable quality across all samples.}
\label{fig:sd_variant_extra}
\end{figure*}

\section{Generative Models and Defenses}
\label{appendix:defense-implementation}

\para{Defense implementations.}

\textit{\univclip{}.} Here, the hyperparameters provided by the authors did not yield the best performance. \uclip{} worked best on the SD dataset when it was fine-tuned for 200 epochs using Adam optimizer with a learning rate of $5e-4$, combined with 10 epochs for early stopping. For the StyleCLIP dataset, we used the same setting with the difference of using a learning rate of $1e-3$. Both settings required data augmentation.

\textit{\defake{}.} Optimal performance was not achieved with author-provided hyperparameters. For the SD dataset, we fine-tuned the model for 200 epochs at a learning rate of $5e-4$. For the StyleCLIP dataset, it was relatively harder to achieve good performance, but fine-tuning for 200 epochs at a learning rate $5e-5$ gives the best results. One of the reasons for struggling with our StyleCLIP dataset may be the low discriminatory value in text features between real and fake data, which is key for classification with \defake{}.

\textit{\dct{}.} Instead of a 64$\times$64 central cutout, we modify the pipeline to extract log(DCT) features from the entire image for a more comprehensive analysis. We also use a smaller version of our dataset, i.e. a balanced dataset of 3$K$, 1$K$, and 2$K$ images for training, validation, and testing, respectively to decrease computational complexity. As it is a simple linear classifier, we train the model from scratch for 10 epochs with the SGD optimizer with a learning rate of $1e-2$ and a weight decay regularization of $1e-3$ for both datasets.

\textit{\patch{}.} Best performance was obtained using the author-provided hyperparameters when fine-tuned on both the datasets with learning rate of $1e-3$ and 100 epochs.

\textit{\texture{}.} Author-provided hyper-parameters did not yield optimal performance for SD dataset, but after fine-tuning for 100 epochs with Adam optimizer, using learning rate of $1e-4$ without weight decay gave optimal performance. For StyleCLIP, fine-tuning for 10 epochs with author-provided hyper parameters gave high performance.

\textit{\resyn{}.} Author-provided hyperparameters did not yield good performance here. The best performance was obtained by fine-tuning on both datasets with a learning rate of $1e-2$ and a reconstruction learning rate of $1e-3$. \resyn{} was trained for 150 and 100 epochs for SD and StyleCLIP datasets, respectively.

\textit{\cnn{}.} We fine-tune both author-provided checkpoints with given hyperparameters and found the best performance.
with the Blur+JPEG (0.5) model on the SD dataset, and Blur+JPEG (0.1) model on the StyleCLIP dataset.

\textit{\meso{}.} A MesoInception-4 model was fine-tuned for 100 epochs using MSE loss with Adam optimizer and a learning rate of $1e-3$ for both datasets.

\para{Other generative models.} We considered the use of other generative models for our evaluation but did not include them for one or more of the following reasons: (1) \textit{Unavailability of training code or pretrained checkpoints}: High-quality generative models require significant computational effort~\cite{2EmadonX9:online}. It is impractical to train them from scratch with limited computational resources. For goal 3 (Section~\ref{sec:background}), we require training code to adversarially update the generator. Both requirements exclude models such as DALL-E~\cite{ramesh2021zero}, GLIDE~\cite{nichol2021glide}, StyleGAN-T~\cite{sauer2023stylegan}, and many others~\cite{kang2023scaling, saharia2022photorealistic}. 
(2) \textit{Poor quality imagery}: We aim to consider a challenging setting where the deepfake images are of high quality; otherwise, they can be easily flagged by human inspection. This criteria excludes models such as CogView~\cite{ding2021cogview} and VQGAN-CLIP~\cite{crowson2022vqgan}. For example, we find that VQGAN-CLIP generates images with repeated artifacts that can be easily flagged.

\section{Details of User-customized SD models}
\label{sec:user-created-model-details}
\noindent Table~\ref{tab:att_custom_lora} shows details of the 16 user-customized models, which includes both LoRA and FM based models.

\begin{table*}[!t]
\centering
\footnotesize
\setlength{\tabcolsep}{3pt}
\setlength\extrarowheight{3pt}
\begin{tabular}{c"c"l|l}
\textbf{Type} & \textbf{\begin{tabular}[c]{@{}c@{}}Model\\ ID\end{tabular}} & \multicolumn{1}{c|}{\textbf{Description}} & \multicolumn{1}{c}{\textbf{Links}} \\ \hlinewd{1.1pt}
\multirow{8}{*}{LoRA} & 1 & Tweak brightness & \href{https://civitai.com/models/70034/brightness-tweaker-lora-lora}{civitai.com/models/70034/brightness-tweaker-lora-lora} \\ 
 & 2 & Add details & \href{https://civitai.com/models/58390/detail-tweaker-lora-lora}{civitai.com/models/58390/detail-tweaker-lora-lora} \\ 
 & 3 & Increase sharpness & \href{https://civitai.com/models/69267/sharpness-tweaker-lora-lora}{civitai.com/models/69267/sharpness-tweaker-lora-lora} \\ 
 & 4 & Enhance image & \href{https://civitai.com/models/78283/elixir-enhancer-lora}{civitai.com/models/78283/elixir-enhancer-lora} \\ 
 & 5 & Improve contrast & \href{https://civitai.com/models/48139/lowra}{civitai.com/models/48139/lowra} \\ 
 & 6 & Add aesthetic details & \href{https://civitai.com/models/82098/add-more-details-detail-enhancer-tweaker-lora}{civitai.com/models/82098/add-more-details-detail-enhancer-tweaker-lora} \\ 
 & 7 & Reduce image noise & \href{https://civitai.com/models/13941?modelVersionId=16576}{civitai.com/models/13941?modelVersionId=16576} \\ 
 & 8 & Tweak skin texture & \href{https://civitai.com/models/134883/skintextureslider-plastic-skin-realistic-skin}{civitai.com/models/134883/skintextureslider-plastic-skin-realistic-skin} \\ \hline
\multirow{8}{*}{FM} & 9 & Increase realism (AbsoluteReality) & \href{https://huggingface.co/Lykon/AbsoluteReality/tree/main}{huggingface.co/Lykon/AbsoluteReality/tree/main} \\ 
 & 10 & Real-life photographs (AnalogDiffusion) & \href{https://civitai.com/models/1265/analog-diffusion}{civitai.com/models/1265/analog-diffusion} \\ 
 & 11 & Photorealism (DreamlikePhotoreal) & \href{https://huggingface.co/dreamlike-art/dreamlike-photoreal-2.0}{huggingface.co/dreamlike-art/dreamlike-photoreal-2.0} \\ 
 & 12 & Artistic and realistic (DreamShaper) & \href{https://civitai.com/models/4384/dreamshaper}{civitai.com/models/4384/dreamshaper} \\ 
 & 13 & High-quality styled images (EpicDiffusion) & \href{https://civitai.com/models/3855/epic-diffusion}{civitai.com/models/3855/epic-diffusion} \\ 
 & 14 & Photorealistic image (epiCRealism) & \href{https://civitai.com/models/25694/epicrealism}{civitai.com/models/25694/epicrealism} \\ 
 & 15 & HDR photography (FotoAssistedDiffusion) & \href{https://huggingface.co/Dunkindont/Foto-Assisted-Diffusion-FAD\_V0}{huggingface.co/Dunkindont/Foto-Assisted-Diffusion-FAD\_V0} \\ 
 & 16 & Realistic artwork (Haveall) & \href{https://civitai.com/models/118799/haveall}{civitai.com/models/118799/haveall}
\end{tabular}%
\caption{Details of LoRA and FM models used in our work.}
\label{tab:att_custom_lora}
\end{table*}

\section{Generating Adversarial Fake Images}
\label{sec:adv-attack-details}

\para{Text prompts used for our attack.} StyleCLIP requires a \textit{neutral} and \textit{target} text. As we only work with human face images, \textit{neutral} text is always ``a face''. The list of \textit{target} text is: (i) a smiling face, (ii) a happy face, (iii) a sad face, (iv) a face with glasses, (v) a face with lipstick, (vi) a face with blue eyes, (vii) a face with brown hair, and (viii) a face with surprise.

\para{Training configuration for surrogates and generator.} For ViT and EfficientNet, we use the SGD optimizer with $1e-3$ learning rate, $0.9$ momentum, and set $\gamma$ and $\delta$ to $0.1$ and $1.0$, respectively. For CLIP-ResNet, we use the same optimizer settings with $\gamma$ and $\delta$ set to $0.02$ and $1.0$, respectively. In all cases, we have set $\alpha$ and $\beta$ to $9.0$ and $0.12$, respectively.

\textit{Fine-tuning Surrogate Classifiers.} We use FFHQ and StyleGAN2 generated images as \textit{real} and \textit{fake} data, and ensure that there is no overlap with the data to fine-tune the defenses, i.e. StyleCLIP dataset. We take 10$K$, 2$K$, and 2$K$ images per class for training, validation, and testing as our \textbf{surrogate StyleCLIP dataset}. As surrogates, we choose ViT as it is the state-of-the-art transformer-based encoder model used for image classification tasks, EfficientNet because of its superior performance despite being 21 times smaller than comparable ConvNets, and CLIP-ResNet because it is pretrained with an Internet-scale dataset. 
To build the surrogates, we choose ViT-base (for ViT), EfficientNet-B0 (for EfficientnET) and 64$\times$ ResNet-50 (for CLIP-ResNet), and add a linear layer for binary classification. We fine-tune ViT, EfficientNet and CLIP-ResNet for 20, 30 and 30 epochs, respectively, with the Adam optimizer at a learning rate of $5e-5$, $5e-4$ and $1e-3$. 

For ViT and EfficientNet, all layers of the models are fine-tuned, whereas for CLIP-ResNet, only the linear layer added at the end is fine-tuned to obtain optimal performance.

\para{Explaining differences between Source fake, Non-adversarial fake and Adversarial Fake.}
We explain the changes that cause fake image to be adversarial against defenses. In Figure~\ref{fig:quality_adv_attack}, we see the following differences between \textit{non-adversarial fake} and \textit{adversarial fake} images after our attack: (i) second image from the left has brighter skin-tone, (ii) third image has green eyes, (iii) fifth image has more texture under the eyes, (iv) sixth image has fewer folds in the skin with a lighter skin tone. Other images are too similar to point anything out. The target text for the images (from left to right) are (i) a face with lipstick, (ii) a face with brown hair, (iii) a face with brown hair, (iv) a face with lipstick, (v) a smiling face, (vi) a smiling face, and (vii) a face with surprise.

\section{Ensemble with \uclip{}}
\label{sec:univclip_ensemble}

\begin{table}[!htb]
\centering
\small
\setlength{\tabcolsep}{3pt}
\setlength\extrarowheight{3pt}
\begin{tabular}{l"c|c|c}
\textbf{Defenses} & \textbf{Precision (\%)} & \textbf{Recall (\%)} & \textbf{F1 (\%)} \\ \hlinewd{1.1pt}
\uclip{} & 90.09 & 68.56 & 76.8 \\ 
\uclip{} + \texture{} & 90.9 & 81.84 & 85.26 \\ 
\uclip{} + \resyn{} & 80.22 & 82.65 & 80.98 \\ 
\uclip{} + \cnn{} & 90.51 & 74.37 & 80.64 \\ 
\uclip{} + \meso{} & 87.27 & 82.73 & 84.35 \\ 
\uclip{} + \defake{} & 86.4 & 81.89 & 83.42
\end{tabular}%
\caption{Improving generalization performance by creating an ensemble model by combining \uclip{} defense with the other 5 defenses. We show the average scores across the 16 SD custom checkpoints.}
\label{tab:univclip_ensemble}
\end{table}

\section{Frequency Spectrum Analysis of User-customized model images}
\label{sec:freqartifact_sd}

\begin{figure}[!htb]
\centering
\includegraphics[scale=0.48]{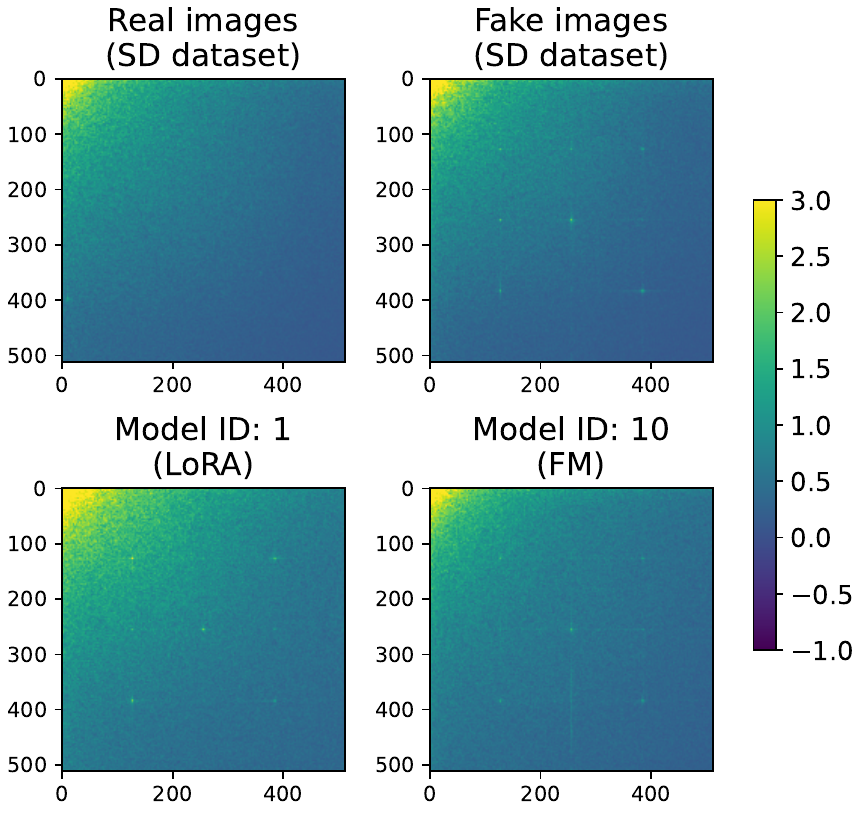}
\caption{Average log (DCT) spectrum calculated over $1K$ images for each category. ``Real" and ``Fake" images are from our SD dataset, and the bottom 2 plots correspond to images from LoRA and FM models in Table~\ref{tab:att_custom_lora}. LoRA and FM images show frequency spectrum artifacts similar to fake images, but they are absent for real images. This explains the strong detection performance of \dct{} over user-customized models.}
\label{fig:freqartifact_sd}
\end{figure}


\newpage 



\section{Meta-Review}
\noindent The following meta-review was prepared by the program committee for the 2024 IEEE Symposium on Security and Privacy (S\&P) as part of the review process as detailed in the call for papers.

\subsection{Summary}
\noindent This paper studies the shortcomings of existing deepfake defense techniques by conducting experiments with two groups of attacks (1) customized generative model and (2) foundation models to generate adversarial deepfakes. The experiments show that current deepfake defense techniques are ineffective against different attack vectors and performance can be heavily impacted. Further, the paper explores different possible ways to improve the deepfake defense.

\subsection{Scientific Contributions}
\begin{itemize} 
\item Independent Confirmation of Important Results with Limited Prior Research 
\item Provides a Valuable Step Forward in an Established Field 
\end{itemize}

\subsection{Reasons for Acceptance}
\begin{enumerate} 
\item This paper identifies major shortcomings of existing deepfake defense techniques. The authors select 8 state-of-the-art deepfake detectors and evaluate them with different adaptive attacks leveraging customized generative models and foundation models to generate adversarial samples. The evaluation shows that the performance of deepfake detectors is heavily impacted against different adversarial attacks and the detection rate can be lowered by over 53\%. This paper provides a valuable step forward in deepfake research. 
\item The number of deepfake images and videos is increasing rapidly with the popularity of open-source AI models and tools. This paper discusses several valuable insights to improve deepfake detection techniques against novel adversarial attacks. \end{enumerate}

\subsection{Noteworthy Concerns}  
\begin{enumerate} 
\item The paper uses the SD dataset for generalizability and the StyleCLIP dataset for adversarial attack evaluation. This might introduce some biases in the evaluation. This paper would benefit from a more generalized approach to the evaluation. 
\end{enumerate}

\section{Response to the Noteworthy Concerns} 
\noindent In Section~\ref{sec:generalization_testing}, we use the SD dataset because among the two models (SD and StyleCLIP), only the SD model has seen widespread user-customization. Therefore, we cannot include a version of the StyleCLIP dataset to study generalization. Recall that the goal of this section is to study the impact of user-customization on deepfake detection. In Section~\ref{sec:advattack}, evaluating our attack using SD can be accomplished using a classifier-guidance approach, which we leave for future work.
We discuss these points in the following places: (1) at the end of the first paragraph in Section~\ref{sec:generalization_testing}, and (2) under Point 5 in Section~\ref{sec:future-work}.

\end{appendices}